\documentclass[prd,superscriptaddress,notitlepage,showpacs,nofootinbib,longbibliography]{revtex4-1}

\usepackage{graphicx}
\usepackage[utf8]{inputenc} 
\usepackage{placeins}
\usepackage{amsmath}
\usepackage{amssymb}
\usepackage{float}
\usepackage{xcolor,color,colortbl}
\usepackage{comment}
\usepackage{multirow}
\usepackage{color}
\usepackage{soul}

\def\beq{\begin{equation}}
\def\eeq{\end{equation}}
\def\bea{\begin{eqnarray}}
\def\eea{\end{eqnarray}}

\usepackage{amsbsy}
\usepackage{bm}
\usepackage{color}
\usepackage{fancyhdr}
\usepackage[pdftex]{epsfig}
\usepackage[colorlinks=true,linkcolor=blue,citecolor=blue,urlcolor=magenta,filecolor=blue]{hyperref}
\usepackage{slashed}

\begin{document}

\bigskip

\vspace{2cm}

\title{Charged current $b \to c \tau \bar{\nu}_\tau$ anomalies in a general $W^\prime$ boson scenario}
\vskip 6ex

\author{John D. G\'{o}mez}
\email{johnd.gomez@udea.edu.co}
\affiliation{Instituto de F\'{i}sica, Universidad de Antioquia, A. A. 1226, Medell\'{i}n, Colombia}
\affiliation{Facultad de Ciencias Exactas y Aplicadas, Instituto Tecnol\'{o}gico Metropolitano, Calle 73 No. 76 A - 354, V\'{i}a el Volador, Medell\'{i}n, Colombia }
\author{N\'{e}stor Quintero}
\email{nestor.quintero01@usc.edu.co}
\affiliation{Facultad de Ciencias B\'{a}sicas, Universidad Santiago de Cali, Campus Pampalinda, Calle 5 No. 62-00, C\'{o}digo Postal 76001, Santiago de Cali, Colombia}
\author{Eduardo Rojas}
\email{eduro4000@gmail.com}
\affiliation{Departamento de Física, Universidad de Nari\~no, A.A. 1175, San Juan de Pasto, Colombia}

\bigskip

\begin{abstract}
Very recent experimental information obtained from the Belle experiment, along with that accumulated by the BABAR and LHCb experiments, has shown the existence of anomalies in the ratios $R(D)$ and $R(D^{*})$ associated with the charged-current transition $b \to c \tau \bar{\nu}_\tau$. Although the Belle measurements are in agreement with the standard model (SM) predictions, the new experimental world averages still exhibit a tension. In addition, the $D^\ast$ longitudinal polarization $F_L(D^\ast)$ related with the channel $B \to D^\ast \tau \bar{\nu}_\tau$ observed by the Belle Collaboration and the ratio $R(J/\psi)$ measured by the LHCb Collaboration also show discrepancies with their corresponding SM estimations. We present a model-independent study based on the most general effective Lagrangian that yields a tree-level effective contribution to the transition $b \to c \tau \bar{\nu}_\tau$ induced by a general $W'$ boson. Instead of considering any specific new physics (NP) realization, we perform an analysis by considering all of the different chiral charges to the charm-bottom and $\tau$-$\nu_{\tau}$ interaction terms with a charged $W'$ boson that explain the anomalies.
We present a phenomenological study of parameter space allowed by the new experimental $b \to c \tau \bar{\nu}_\tau$ data and with the mono-tau signature $pp \to \tau_h X + \rm{MET}$ at the LHC. For comparison we include some of the $W'$ boson NP realizations that have already been studied in the literature.
\end{abstract}

\maketitle

\section{Introduction}

The $B$ meson system has constituted a good scenario for studying, on both theoretical and experimental levels, the Standard Model (SM) as well as for exploring new physics (NP) effects at low-energy scales. Particularly, semileptonic and leptonic $B$ meson decays offer an excellent place to test lepton universality (LU), so far one of the cornerstones of the SM. Any mismatch between the theoretical and experimental predictions may be an indication of LU violation, and therefore a hint of NP beyond the SM~\cite{Ciezarek:2017yzh,Bifani:2018zmi}.

\begin{figure*}[!b]
\centering
\includegraphics[scale=0.5]{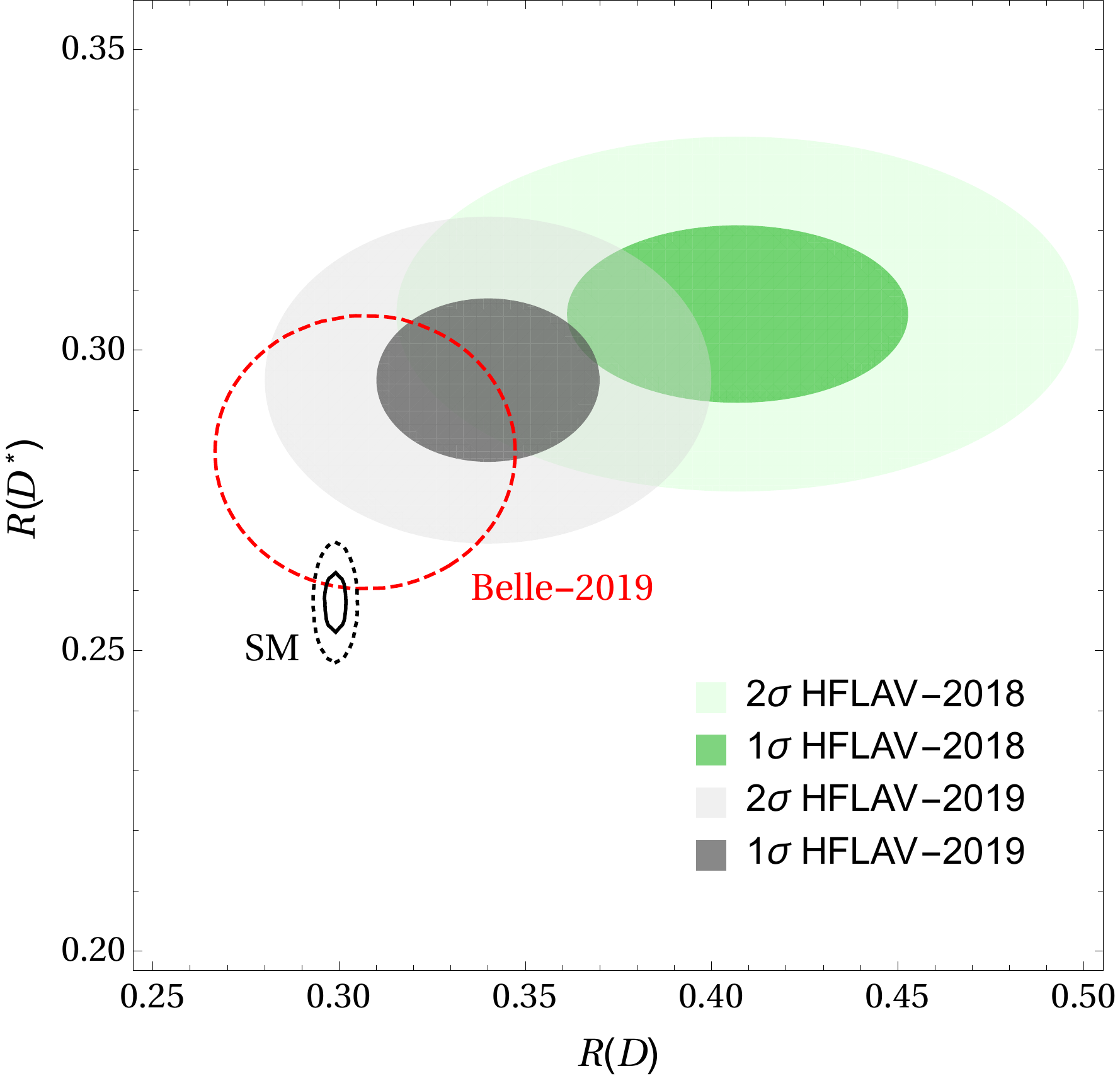}
\caption{\small The HFLAV-2018 and HFLAV-2019 averages (green and gray regions, respectively)~\cite{Amhis:2016xyh,HFLAVsummer} in the $R(D)$ vs $R(D^\ast)$ plane. The black ($1\sigma$ solid and  $2\sigma$ dotted) and red (dashed) contours shows the SM predictions and the recent Belle measurements~\cite{Abdesselam:2019dgh}, respectively.}
\label{Fig1}
\end{figure*}

The BABAR Collaboration in 2012 was the first experiment that reported a disagreement on the measurements of the ratio of semileptonic $B$ decays ($b \to c$ transition processes)~\cite{Lees:2012xj,Lees:2013uzd} 
\begin{equation} \label{R_D}
R(D^{(\ast)}) = \dfrac{{\rm BR}(B \to D^{(\ast)}\tau \bar{\nu}_\tau)}{{\rm BR}(B \to D^{(\ast)}\ell^\prime \bar{\nu}_{\ell^\prime})} , \ \ \ell^\prime = e \ {\rm or} \ \mu ,
\end{equation}

\noindent compared with the SM predictions~\cite{Fajfer:2012vx,Fajfer:2012jt,Bailey:2012jg}. These discrepancies were later confirmed by Belle~\cite{Huschle:2015rga,Sato:2016svk,Hirose:2017dxl,Hirose:2016wfn}, and LHCb~\cite{Aaij:2015yra,Aaij:2017deq,Aaij:2017uff} experiments by means of different techniques. 
Theoretical progress on the SM calculations of $R(D^{(\ast)})$ has been made recently~\cite{Bigi:2016mdz,Aoki:2016frl,Bernlochner:2017jka,Jaiswal:2017rve,Bigi:2017jbd}, with average values~\cite{Amhis:2016xyh,HFLAVsummer} shown in Table~\ref{Table1}. Despite all of these advancements, the experimental measurements on $R(D^{(\ast)})$ still exhibit a deviation from the SM expectations. Nevertheless, things seem to have changed, and the tension has been reduced with the new results on $R(D^{(\ast)})$ that the Belle Collaboration has recently released~\cite{Abdesselam:2019dgh} (as presented in Table~\ref{Table1}), which are now in agreement with the SM predictions within 0.2$\sigma$ and 1.1$\sigma$, respectively. Incorporating these Belle results, in Table~\ref{Table1} we display the new 2019 world averages values reported by the Heavy Flavor Averaging Group (HFLAV) on the measurements of $R(D)$ and $R(D^*)$~\cite{Amhis:2016xyh,HFLAVsummer}, which now exceed the SM predictions by 1.4$\sigma$ and 2.5$\sigma$, respectively. To see the incidence of the very recent Belle results, in Figure~\ref{Fig1}, we plot the $R(D)$ vs $R(D^\ast)$ plane by showing the  
HFLAV-2018 average (green region) and the new HFLAV-2019 average (blue region)~\cite{Amhis:2016xyh,HFLAVsummer}, at both $1\sigma$ and $2\sigma$. The black (solid $1\sigma$ and dotted $2\sigma$) and red (dashed) contours show the SM predictions and the recent Belle measurements, respectively. This $R(D)$ vs $R(D^\ast)$ plot illustrates how the anomalies have been significantly narrowed due to the new Belle data.


Further hints of lepton flavor universality violation in the charged current  $b \to c  \tau \bar{\nu}_\tau$ have recently been obtained by LHCb in the measurement of the ratio~\cite{Aaij:2017tyk} 
\beq \label{R_Jpsi}
R(J/\psi) = \dfrac{{\rm BR}(B_c \to J/\psi\tau\bar{\nu}_\tau)}{{\rm BR}(B_c \to J/\psi\mu \bar{\nu}_\mu)},
\eeq

\noindent which also shows tension with regard to the SM prediction (around 2$\sigma$)~\cite{Dutta:2017xmj,Watanabe:2017mip,Murphy:2018sqg,Cohen:2018dgz,Issadykov:2018myx,Azizi:2019aaf}. In further calculations, we will use the theoretical prediction of Ref.~\cite{Watanabe:2017mip} (see Table~\ref{Table1}), which is in agreement with other estimations~\cite{Dutta:2017xmj,Murphy:2018sqg,Cohen:2018dgz,Issadykov:2018myx,Azizi:2019aaf}. 
Additionally, polarization observables associated with the channel $B \to D^\ast \tau \bar{\nu}_\tau$ have been observed in the Belle experiment, namely, the $\tau$ lepton polarization $P_\tau(D^\ast)$~\cite{Hirose:2017dxl,Hirose:2016wfn} and the $D^\ast$ longitudinal polarization $F_L(D^\ast)$~\cite{Abdesselam:2019wbt}. We present in Table~\ref{Table1} these measurements, as well as their corresponding SM values~\cite{Tanaka:2012nw,Alok:2016qyh}, which also exhibit a deviation from the experimental data.

\begin{table}[!t]
\centering
\renewcommand{\arraystretch}{1.2}
\renewcommand{\arrayrulewidth}{0.8pt}
\caption{\small Experimental status on observables related to the charged transition $b \to c \tau \bar{\nu}_\tau$.}
\begin{tabular}{ccc}
\hline
Observable & Exp measurement & SM prediction \\
\hline
$R(D)$ & $0.307 \pm 0.037 \pm 0.016$ \ Belle-2019~\cite{Abdesselam:2019dgh} & 0.299 $\pm$ 0.003~\cite{Amhis:2016xyh,HFLAVsummer} \\
       & $0.340 \pm 0.027 \pm 0.013$ \ HFLAV~\cite{Amhis:2016xyh} & \\
$R(D^\ast)$ & $0.283 \pm 0.018 \pm 0.014$ \ Belle-2019~\cite{Abdesselam:2019dgh} & 0.258 $\pm$ 0.005~\cite{Amhis:2016xyh,HFLAVsummer} \\
        & $0.295 \pm 0.011 \pm 0.008$ HFLAV~\cite{Amhis:2016xyh} & \\
$R(J/\psi)$ & $0.71 \pm 0.17 \pm 0.18$~\cite{Aaij:2017tyk} & 0.283 $\pm$ 0.048~\cite{Watanabe:2017mip}  \\
$P_\tau(D^\ast)$ & $- 0.38 \pm 0.51 ^{+0.21}_{-0.16}$~\cite{Hirose:2017dxl,Hirose:2016wfn} &  $-0.497 \pm 0.013$~\cite{Tanaka:2012nw} \\
$F_L(D^\ast)$ & $0.60 \pm 0.08 \pm 0.035$~\cite{Abdesselam:2019wbt} & $0.46 \pm 0.04$~\cite{Alok:2016qyh}  \\
$R(X_c)$ & 0.223 $\pm$ 0.030~\cite{Kamali:2018bdp} & 0.216 $\pm$ 0.003~\cite{Kamali:2018bdp} \\
\hline
\end{tabular} \label{Table1}
\end{table}


The incompatibility of these measurements with the SM could be an evidence of LU violation in $B$ decays and, therefore, an indication of NP sensitive to the third generation of leptons. In order to understand these discrepancies, an enormous number of theoretical studies have been proposed. On the one hand, model-independent analyses of the impact of NP effective operators have been extensively studied (for the most recent ones that include the new Belle measurements, see Refs.~\cite{Murgui:2019czp,Bardhan:2019ljo,Shi:2019gxi,Asadi:2019xrc,Blanke:2019qrx})\footnote{For previous works, see, for instance, Refs.~\cite{Watanabe:2017mip,Alok:2017qsi,Huang:2018nnq,Azatov:2018knx,Bhattacharya:2018kig,Jung:2018lfu,Tran:2018kuv,Biswas:2018jun,Iguro:2018vqb,Blanke:2018yud}.}. On the other hand, particular NP scenarios such as charged scalars~\cite{Biswas:2018jun,Iguro:2018fni,Fraser:2018aqj,Martinez:2018ynq,Dhargyal:2016eri,Crivellin:2015hha,Iguro:2017ysu,Wei:2017ago,Celis:2016azn,Chen:2017eby,Sakaki:2012ft,Crivellin:2012ye,Crivellin:2013wna,Celis:2012dk,Ko:2012sv,Ko:2017lzd,Li:2018rax}, leptoquarks (both scalar and vector)~\cite{Hati:2019ufv,Hati:2018fzc,Assad:2017iib,Fornal:2018dqn,Yan:2019hpm,Cornella:2019hct,Becirevic:2016yqi,Becirevic:2018afm,Alonso:2015sja,Calibbi:2015kma,Fajfer:2015ycq,Barbieri:2015yvd,Barbieri:2016las,Hiller:2016kry,Bhattacharya:2016mcc,Buttazzo:2017ixm,Kumar:2018kmr,Assad:2017iib,DiLuzio:2017vat,Calibbi:2017qbu,Barbieri:2017tuq,Blanke:2018sro,Greljo:2018tuh,Bauer:2015knc,Bordone:2018nbg,Crivellin:2018yvo,DiLuzio:2018zxy,Cai:2017wry,Crivellin:2017zlb,Li:2016vvp,Das:2016vkr,Faroughy:2016osc,Sahoo:2016pet,Chen:2017hir,Sakaki:2013bfa}, extra gauge bosons~\cite{Iguro:2018fni,Dasgupta:2018nzt,He:2012zp,He:2017bft,Boucenna:2016qad,Boucenna:2016wpr,Greljo:2015mma,Faroughy:2016osc, Abdullah:2018ets,Greljo:2018tzh,Carena:2018cow,Babu:2018vrl,Asadi:2018wea,Greljo:2018ogz,Robinson:2018gza,Asadi:2018sym}, right-handed neutrinos~\cite{Li:2018rax,Carena:2018cow,Babu:2018vrl,Asadi:2018wea,Greljo:2018ogz,Robinson:2018gza,Asadi:2018sym,Azatov:2018kzb,Cvetic:2017gkt}, R-parity violating supersymmetric couplings~\cite{Tanaka:2012nw,Altmannshofer:2017poe,Deshpande:2016cpw,Deshpande:2012rr,Zhu:2016xdg,Wei:2018vmk,Hu:2018lmk,Trifinopoulos:2018rna}; have been studied as well. 
Complementary tests at the LHC searches of some of these scenarios have been also explored~\cite{Iguro:2018fni,Altmannshofer:2017poe,Iguro:2017ysu,Abdullah:2018ets,Iguro:2018vqb,Greljo:2018tzh,Greljo:2018ogz}. Furthermore, the polarizations of the $\tau$ lepton and $D^\ast$ are also useful observables to potentially distinguish the underlying NP~\cite{Tanaka:2012nw,Alok:2016qyh,Iguro:2018vqb,Blanke:2018yud}.

The potential NP scenarios that could explain the $R(D^{(*)})$ and $R(J/\psi)$ anomalies would also affect the branching ratio associated with the leptonic decay $B_c^{-} \to \tau^{-} \bar{\nu}_\tau$~\cite{Alonso:2016oyd,Akeroyd:2017mhr} since all of them are generated by the same quark level transition $b \to c  \tau \bar{\nu}_\tau$. In Ref.~\cite{Alonso:2016oyd}, a constraint of ${\rm BR}(B_c^{-} \to \tau^{-} \bar{\nu}_\tau) \lesssim 30 \%$ is imposed by considering the lifetime of $B_c$, while a stronger bound of ${\rm BR}(B_c^{-} \to \tau^{-} \bar{\nu}_\tau) \lesssim 10 \%$ has been obtained in Ref.~\cite{Akeroyd:2017mhr} from the LEP data taken at the $Z$ peak.  In the SM, the branching fraction of this tauonic decay is given by the expression~\cite{Alonso:2016oyd,Akeroyd:2017mhr}
\bea \label{SM_leptonic}
{\rm BR}(B_c^-  \to  \tau^- \bar{\nu}_\tau)_{\rm SM} &=&\tau_{B_c} \dfrac{G_F^2}{8\pi} |V_{cb}|^2 f_{B_c}^2 m_{B_c} m_{\tau}^2  \Big(1- \dfrac{m_{\tau}^2}{m_{B_c}^2}\Big)^2 ,
\eea
\noindent where $G_F$ is the Fermi constant, $V_{cb}$ denotes the Cabbibo-Kobayashi-Maskawa (CKM) matrix element involved, and $f_{B_c}$ and $\tau_{B_c}$ are the $B_c^-$ meson decay constant and lifetime, respectively. By using the following input values, $\tau_{B_c} = (0.507 \pm 0.009)$ ps, $m_{B_c} = 6.2749$ GeV, and $|V_{cb}| = (40.5 \pm 1.5) \times 10^{-3}$ from the Particle Data Group (PDG)~\cite{Tanabashi:2018oca} and $f_{B_c} = (434 \pm 15)$ MeV from lattice QCD~\cite{Colquhoun:2015oha}, we get a value of
\begin{equation} \label{tauonicBcdecay_SM}
{\rm BR}(B_c^- \to \tau^- \bar{\nu}_\tau)_{\rm SM} =  (2.16 \pm 0.16) \%. 
\end{equation}

\noindent It is worth mentioning that by taking the value for $|V_{cb}| = (39.18 \pm 0.94 \pm 0.36) \times 10^{-3}$ reported by HFLAV~\cite{Amhis:2016xyh}, a value of ${\rm BR}(B_c^- \to \tau^- \bar{\nu}_\tau)_{\rm SM} = (2.02 \pm 0.11) \%$ is obtained, which is consistent with Eq.~\eqref{tauonicBcdecay_SM}. For later use in our phenomenological analysis, we will take Eq.~\eqref{tauonicBcdecay_SM} and the upper limit ${\rm BR}(B_c^{-} \to \tau^{-} \bar{\nu}_\tau) \lesssim 10 \%$. Moreover, we will consider the inclusive semileptonic decay $B \to X_c \tau^-\bar{\nu}_\tau$ that is generated via the same transition $b \to c  \tau^- \bar{\nu}_\tau$. Including non-perturbative corrections of the order $\mathcal{O}(1/m_b^2)$ and using the $1S$ mass scheme, in Ref.~\cite{Kamali:2018bdp}, a very recent estimation has been calculated, $R(X_c)_{\rm SM}= 0.228 \pm 0.030$, that is in agreement ($0.2 \sigma$) with the experimental value $R(X_c)_{\rm exp} = 0.223 \pm 0.030$~\cite{Kamali:2018bdp} (these values are also collected in Table~\ref{Table1}). 

In light of the new HFLAV world average values $R(D^{(\ast)})$~\cite{Amhis:2016xyh,HFLAVsummer} (due to the very recent Belle measurements~\cite{Abdesselam:2019dgh}) and the polarization observables $P_\tau(D^\ast)$, $F_L(D^\ast)$ measured by Belle~\cite{Hirose:2017dxl,Hirose:2016wfn,Abdesselam:2019wbt}, in this work we look into the interpretation of these charged-current $B$ anomalies driven by a general $W^\prime$ gauge boson scenario. Without invoking any particular NP model, we provide a model-independent study based on the most general effective Lagrangian given in terms of the flavor-dependent couplings  $\epsilon_{cb}^{L,R}$ and $\epsilon_{\tau\nu_\tau}^{L,R}$ of the currents $(\bar{c} \gamma_\mu P_{L,R} b)$ and $(\bar{\tau} \gamma^\mu P_{L,R} \nu_{\tau})$, respectively (see Sec.~\ref{model} for details), whichyields a tree-level effective contribution to the $b \to c \tau \bar{\nu}_\tau$ transition. We implement a $\chi^2$ analysis by considering all of the scenarios with different chiral charges that explain the $R(D^{(\ast)})$ discrepancies. 
We also analyze the effect of taking into account all of the charged transition $b \to c \tau \bar{\nu}_\tau$ observables namely $R(J/\psi)$, $P_\tau(D^\ast), F_L(D^\ast), R(X_c)$, and BR$(B_c^{-} \to \tau^{-} \bar{\nu}_\tau)$. We present a phenomenological analysis of parameter space allowed by the experimental data, and for comparison, we include some of the $W'$ boson NP realizations that have already been studied in the literature~\cite{Greljo:2015mma,Faroughy:2016osc,Abdullah:2018ets,Greljo:2018tzh,Asadi:2018wea,Greljo:2018ogz,He:2012zp,He:2017bft,Babu:2018vrl}. Most of these models were implemented by considering the previous HFLAV averages and, in addition, not all of them considered the polarization observables $P_\tau(D^\ast)$ and $F_L(D^\ast)$; therefore, we explore which of these benchmark models are still favored (or disfavored) by the new $b \to c \tau \bar{\nu}_\tau$ data.

It is important to remark that since we are not implementing any NP realizations in our analysis, we will get out of our discussion the possible connection with a $Z^\prime$ boson that appears in particular UV completions, as done, for instance, in Refs.~\cite{He:2017bft,Boucenna:2016qad,Greljo:2015mma,Faroughy:2016osc,Greljo:2018tzh,Asadi:2018wea,Greljo:2018ogz,Robinson:2018gza,Asadi:2018sym}.

This work is organized as follows. In Sec.~\ref{model}, we briefly present the most general charged-current effective Lagrangian for a general $W^\prime$ gauge boson; then, we study its tree-level effective contribution to the observables associated with the semileptonic transition $b \to c \tau \bar{\nu}_{\tau}$. In order to provide an explanation to the charged-current $B$ anomalies, in Sec.~\ref{analysis} we study different parametric models that depend on the choices of the chiral charges and carry out a $\chi^2$ analysis to get the best candidates to adjust the experimental data. Based on this analysis, we explore the two parametric model to determine the regions in parameter space favored by two different datasets, $R(D)$ and $R(D^\ast)$, and all of the $b \to c\tau \bar{\nu}_\tau$ observables, and we make a comparison with some benchmark models studied in the literature. Our main conclusions are given in Sec.~\ref{Conclusion}.


\section{A General $W^\prime$ boson scenario} \label{model}
 
The most general Lorentz invariant Lagrangian describing the couplings of a general $W^\prime$ boson to quarks and leptons may be written as\footnote{See the review \textit{$W^\prime$-boson searches} in the PDG~\cite{Tanabashi:2018oca}.}
\begin{equation} \label{Lag_Wprime}
\mathcal{L}_{\rm eff}^{W^\prime} = \frac{W^\prime_\mu}{\sqrt{2}}  \Big[\bar{u}_i (\epsilon^L_{u_id_j} P_L + \epsilon^R_{u_id_j} P_R)\gamma^\mu d_j + \bar{\ell}_i (\epsilon^L_{\ell_i\nu_{j}} P_L + \epsilon^R_{\ell_i\nu_{j}} P_R)\gamma^\mu \nu_{j}\Big] + {\rm H.c.},
\end{equation} 

\noindent where $P_{R/L} = (1 \pm \gamma_5)/2$ are the right-handed (RH) and left-handed (LH) chirality projectors, respectively; and the coefficients $\epsilon^L_{u_id_j}$, $\epsilon^R_{u_id_j}$, $\epsilon^L_{\ell_i\nu_{j}}$, and $\epsilon^R_{\ell_i\nu_{j}}$ are arbitrary dimensionless parameters that codify the NP flavor effects, with $u_i \in (u,c,t)$, $d_j \in (d,s,b)$ and $\ell_i, \ell_j \in (e,\mu,\tau)$. For simplicity, we consider leptonic flavor-diagonal interactions ($i=j$).  In the SM, only the LH couplings $\epsilon^L_{u_id_j} = g_L V_{u_id_j}$ and $\epsilon^L_{\ell_i \nu_i}= g_L$ are present, with $g_L$ being the $SU(2)_L$ gauge coupling constant, and $V_{u_id_j}$ the corresponding CKM quark matrix element.


In the SM framework, the $b \to c \tau \bar{\nu}_{\tau}$ quark level processes are mediated by a virtual $W$ boson exchange, which is described by the effective Lagrangian
\beq \label{H_eff}
- \mathcal{L}_{\rm eff}(b \to c \tau \bar{\nu}_{\tau})_{\rm SM} = \frac{4 G_F}{\sqrt{2}} V_{cb} (\bar{c} \gamma_\mu P_L b) (\bar{\tau} \gamma^\mu P_L \nu_{\tau}) ,
\eeq

\noindent where $G_F$ is the Fermi coupling constant and $V_{cb}$ is the associated CKM matrix element. According to Eq.~\eqref{Lag_Wprime}, a general $W^\prime$ boson exchange leads to additional tree-level effective interactions to the $b \to c \tau \bar{\nu}_{\tau}$ transition; thus, the total low-energy effective Lagrangian has the following form,
\bea\label{Leff_Total}
- \mathcal{L}_{\rm eff}(b \to c \tau \bar{\nu}_{\tau})_{\rm SM+W^\prime} &=& \frac{4 G_F}{\sqrt{2}} V_{cb}	 \Big[(1+ C_V^{LL} )(\bar{c} \gamma_\mu P_L b) (\bar{\tau} \gamma^\mu P_L \nu_{\tau}) + C_{V}^{RL}(\bar{c} \gamma_\mu P_R b) (\bar{\tau} \gamma^\mu P_L \nu_{\tau}) \nonumber \\ 
&& + C_{V}^{LR}(\bar{c} \gamma_\mu P_L b) (\bar{\tau} \gamma^\mu P_R \nu_{\tau}) +C_{V}^{RR}(\bar{c} \gamma_\mu P_R b) (\bar{\tau} \gamma^\mu P_R \nu_{\tau})  \Big], 
\eea
\noindent where $C^{LL}_{V}, C^{RL}_V, C^{LR}_V$ and $C^{RR}_V$ are the Wilson coefficients associated with the NP operators, particularly the LH and RH vector operator contributions, respectively. These Wilson coefficients depend on the choices of the chiral charges and are defined as  
\bea \label{CLL}
C^{LL}_{V} &\equiv & \dfrac{\sqrt{2}}{4G_F V_{cb}} \dfrac{\epsilon_{cb}^{L} \epsilon_{\tau \nu_\tau}^{L}}{M_{W^\prime}^2}, \label{CLL} \\
C^{RL}_{V} &\equiv & \dfrac{\sqrt{2}}{4G_F V_{cb}} \label{CRL} \dfrac{\epsilon_{cb}^{R} \epsilon_{\tau\nu_\tau}^{L}}{M_{W^\prime}^2}, \label{CRL} \\
C^{LR}_{V} &\equiv & \dfrac{\sqrt{2}}{4G_F V_{cb}} \label{CLR}\dfrac{\epsilon_{cb}^{L} \epsilon_{\tau \nu_\tau}^{R}}{M_{W^\prime}^2}, \label{CLR} \\
C^{RR}_{V} &\equiv & \dfrac{\sqrt{2}}{4G_F V_{cb}} \label{CRR} \dfrac{\epsilon_{cb}^{R} \epsilon_{\tau\nu_\tau}^{R}}{M_{W^\prime}^2}, \label{CRR}
\eea

\noindent with $M_{W^\prime}$ being the ${W^\prime}$ boson mass, and $\epsilon_{cb}^{L,R}$ and $\epsilon_{\tau\nu_\tau}^{L,R}$ the effective flavor-dependent couplings given in Eq.~\eqref{Lag_Wprime}. To accommodate the $b \to c \tau \bar{\nu}_{\tau}$ anomalies, we will adopt the phenomenological assumption in which the $W^\prime$ boson couples only to the bottom-charm quarks and the 	third generation of leptons, i.e., the effective couplings $\epsilon_{cb}^{L,R} \neq 0$ and $\epsilon_{\tau\nu_\tau}^{L,R} \neq 0$ are other from zero, while the other ones are taken to be zero. Therefore, NP effects are negligible for light lepton modes ($e$ or $\mu$), $\epsilon_{e\nu_e}^{L,R}=\epsilon_{\mu\nu_\mu}^{L,R}=0$. For simplicity, we take these effective couplings to be real. 
These are the minimal assumptions in order to provide an explanation to the discrepancies and not to get in conflict or tension with another low-energy LU test; for instance, there is no problem with LU constraints from the bottom-charm loop (mediated by a $W^\prime$ boson) contribution to the $\tau$ lepton decay $\tau \to \ell \nu_\tau\nu_\ell$~\cite{Feruglio:2017rjo}.

As a final remark, let us notice that such a flavor texture assumption for the gauge interactions can be obtained by an approximated $U(2)_q \times U(2)_\ell$ flavor symmetry~\cite{Greljo:2015mma,Faroughy:2016osc} or in SM gauge extensions with additional exotic fermions~\cite{Boucenna:2016qad}.


\subsection{Contribution to the charged-current $b \to c \tau \bar{\nu}_{\tau}$ observables}

According to the above effective Lagrangian \eqref{Leff_Total}, a general $W^\prime$ charged boson exchange will modify the observables associated with the semileptonic transition $b \to c \tau \bar{\nu}_{\tau}$.
The ratios $R(M)$ ($M=D,D^\ast,J/\psi$), and the $D^\ast$ and $\tau$ longitudinal polarizations can be parametrized in terms of  the effective Wilson coefficients $C_V^{LL}$, $C_V^{RL}$, $C_V^{LR}$, and $C_V^{RR}$ as follows~\cite{Iguro:2018vqb,Asadi:2018wea,Asadi:2018sym}
\begin{eqnarray}
R(D) &=& R(D)_{\rm SM} \Big(\big|1+C_V^{LL} + C_V^{RL}\big|^2+\big|C_V^{LR} +C_V^{RR}\big|^2\Big),  \label{RD_RPV} \\ 
R(D^*) &=&   R(D^*)_{\rm SM} \Big(|1 + C_V^{LL}|^2 + |C_V^{RL}|^2+ |C_V^{LR}|^2+ |C_V^{RR}|^2 - 1.81 \ {\rm Re}\big[(1+C_V^{LL})C_V^{RL\ast}+(C_V^{RR}) C_V^{LR\ast}\big] \Big), \label{RDstar_RPV}  \\
R(J/\psi) &=&  R(J/\psi)_{\text{SM}}\Big(|1 + C_V^{LL}|^2 + |C_V^{RL}|^2 + |C_V^{LR}|^2+ |C_V^{RR}|^2- 1.92 \ {\rm Re}\big[(1+C_V^{LL})C_V^{RL\ast}+(C_V^{RR})C_V^{LR\ast}\big]\Big), \label{RJPSI_RPV} \\
F_L(D^*) &=&  F_L(D^*)_{\rm SM} \ r_{D^\ast}^{-1}  \Big( \big|1 + C_V^{LL}-C_V^{RL}\big|^2 + \big|C_V^{RR}-C_V^{LR}\big|^2 \Big),\label{FLD_RPV} \\
P_\tau(D^*) &=&  P_\tau(D^*)_{\rm SM} \ r_{D^\ast}^{-1} \Big(|1 + C_V^{LL}|^2 + |C_V^{RL}|^2-|C_V^{RR}|^2-|C_V^{LR}|^2 - 1.77 \ {\rm Re}\big[(1+C_V^{LL})C_V^{RL\ast} \nonumber \\
&& \ -(C_V^{RR})C_V^{LR\ast}\big]\Big),
\label{PTAU_RPV}
\end{eqnarray}

\noindent respectively, with $r_{D^\ast} = R(D^*) / R(D^*)_{\rm SM}$. The numerical formula for $R(J/\psi)$ have been obtained by using the analytic expressions and form factors given in Ref.~\cite{Watanabe:2017mip}. Similarly, the leptonic decay $B_c^- \to \tau^- \bar{\nu}_{\tau}$ will also be modified~\cite{Iguro:2018vqb,Asadi:2018sym}
\begin{eqnarray} \label{BRBc_modified}
{\rm BR}(B_c^- \to \tau^- \bar{\nu}_{\tau}) &=& {\rm BR}(B_c^- \to \tau^- \bar{\nu}_{\tau})_{\text{SM}} \ \Big(\big|1+C_V^{LL} - C_V^{RL}\big|^2+|C_V^{RR}-C_V^{LR}|^2\Big),
\end{eqnarray}
\noindent as well as the ratio $R(X_c)$ of inclusive semileptonic $B$ decays\footnote{We thank Saeed Kamali for the useful conversations.}~\cite{Kamali:2018bdp}
\begin{eqnarray}
R(X_c) &=& R(X_c)_{\rm SM} \Big( 1+ 1.147 \ \Big[\big|C_V^{LL}\big|^2 + \big|C_V^{RR}\big|^2+ 2{\rm Re}(C_V^{LL}) + \big|C_V^{LR}\big|^2+ \big|C_V^{RL}\big|^2\Big] \nonumber \\
&& \ \ - 0.714 \ {\rm Re}\big[(1+C_V^{LL})C_V^{RL\ast}+(C_V^{RR}) C_V^{LR\ast} \big]\Big). \label{RXc}   
\end{eqnarray}

\noindent In the next section, we will pay attention to the Wilson coefficients $C_V^{LL}$, $C_V^{RL}$, $C_V^{LR}$, and $C_V^{RR}$, given in terms of the effective couplings $\epsilon_{cb}^{L,R}$ and $\epsilon_{\tau\nu_\tau}^{L,R}$ and  the ${W^\prime}$ boson mass, which can provide an explanation for the $b \to c \tau \bar{\nu}_{\tau}$ anomalies.

\section{Phenomenological analysis} \label{analysis}

Table~\ref{Table1} shows the most recent measurements for several flavor observables. In what follows, we will denote these values as $\mathcal{O}_{\text{exp}}$, and the corresponding theoretical expressions $\mathcal{O}_{\text{th}}$ are shown in Eqs.~\eqref{RD_RPV}-\eqref{RXc}.  
When there are no correlations or they are negligible,  the  $\chi^2$ function is given by  the sum of the squared pulls, i.e., $\chi^2=\sum_i \text{pull}_i^2$, where $\text{pull}_i=(\mathcal{O}_\text{exp}^i-\mathcal{O}_\text{th}^i)/\sqrt{\sigma_{\text{exp}}^{i 2}+\sigma_{\text{th}}^{i 2}}$.  Here $\sigma_{\text{exp,th}}^{i}$ corresponds to the experimental~(theoretical) error.
{In order to account for the $R(D)$ and  $R(D^\ast)$ correlation, the contribution of these observables to  the $\chi^2$ function 
should be written as 
\begin{align}
\chi^2_{R(D)\text{-} R(D^\ast)}= \frac{\text{pull}(D) +\text{pull}(D^*)^2-2\rho ~\text{pull}(D)\text{pull}(D^*) }{\sqrt{1-\rho^2}}    
\end{align}
where $\rho= -0.203$ is the $R(D)$-$R(D^\ast)$ correlation reported by HFLAV~\cite{Amhis:2016xyh,HFLAVsummer}.  This effect is important since the  experimental methodology and the theoretical expressions  are quite similar for these observables. 
This correlation does not significantly modify the best-fit point for all models presented here. We neglected the remaining correlations.
From Eq.~\eqref{Lag_Wprime}, it is possible to obtain several models by turning on some of the couplings, while the remaining ones are set equal to zero.  In order  to adjust the  experimental anomalies, any model  must contain a charm-bottom  interaction term  in the quark sector and the corresponding  $\tau$-$\nu_{\tau}$ interaction term in the lepton sector; this means that it is necessary to have at least  two nonzero couplings in the Lagrangian~\eqref{Lag_Wprime}.
These models will be referred to as 2P models, and the corresponding models with three and four nonzero couplings will be referred to as 3P and 4P, respectively.      
Depending on the choices for the chiral charges $(\epsilon^{L,R}_{cb},\epsilon^{L,R}_{\tau\nu_\tau})$, there are four different  2P models, $LL$, $LR$, $RL$, and $RR$. As we will see in the next section,  two of them ($LL$ and $RR$) have already been studied in the literature; however, the  $LR$ and $RL$ models, as far as we know, have not been reported on yet. The same is true for the 3P and 4P models.

In order to check whether it is possible to adjust the deviations of the standard model predictions in these models, we carried out a $\chi^2$ analysis with the seven experimental observables mentioned above. Owing to the absence of the experimental measurement on $B_c^- \to \tau^- \bar{\nu}_{\tau}$, we used the SM estimation given in Eq.~\eqref{tauonicBcdecay_SM}, which is consistent with the strongest upper limit~${\rm BR}(B_c^{-} \to \tau^{-} \bar{\nu}_\tau) < 10 \%$~\cite{Akeroyd:2017mhr}. In Table~\ref{Table2}, we display the SM pulls of all of the $b \to c\tau \bar{\nu}_\tau$ observables. The fit results are shown in Table~\ref{tab:pulls}. In this fit,  the number of degrees of freedom is given by $\text{dof}=7-p$, where $p$ is the number of parameters. The goodness of fit $\chi^2_{\text{min}}/\text{dof}$ is of order 1 for 2P models~(except for the RL model); for the 3P and the 4P models, $\chi^2_{\text{min}}/\text{dof}~\sim 1.4$ and $1.8$, respectively. So, the 2P models represent the best candidates to adjust the experimental anomalies. 
It is important to note that the observables that generate more tension are $R(J/\psi)$ and $F_L(D^*)$, even though these experiments have large uncertainties.
By comparing  Tables~\ref{Table2} and~\ref{tab:pulls}, we can see that,  with respect to the SM, the models with an additional $W^\prime$  decrease the pulls for $R(D)$ and $R(D^\ast)$ without increasing the pulls of the other observables.
In order to keep the couplings in the perturbative regime, we took the mass of the $W^\prime$ boson as $M_{W^\prime}= 1$ TeV.  There is no tension with the current LHC constraints for the $M_{W^\prime}$ (which are above 4~TeV) since we are assuming zero couplings to the SM fermions of the first family. 

\begin{table}[!t]
\begin{center} \textcolor{blue}{
\begin{tabular}{|c|c|c|c|c|c|}
\hline
  $R(D)$  &  $R(D^{\ast})$  &  $R(J/\psi)$   &  $P_\tau(D^{\ast})$  &  $F_L(D^{\ast})$  &  $R(X_c)$ \\
\hline
  1.36    &   2.55          &    1.69        &     0.21             &       1.46        &   0.23\\
\hline
\end{tabular}
\caption{SM pulls of the all $b \to c\tau \bar{\nu}_\tau$ observables}
\label{Table2}}
\end{center}
\end{table}

\begin{table}[!t]
\begin{center}
\begin{tabular}{|c|c|c|c|c|c|c|c|c|c|c|c|c|c|}
\hline
&                          & \multicolumn{7}{| c| }{$\text{pull}_i$} &     &\multicolumn{4}{| c| }{best-fit point}     \\ \hline
& Parameters on&  $R(D)$   & $R(D^{\ast})$  & $R(J/\psi)$    & $P_\tau(D^{\ast})$ & $F_L(D^{\ast})$ & $R(X_c)$ & BR$(B_c \to \tau \bar{\nu})$  & $\chi^2_{\text{min}}$ & $\epsilon_{cb}^L$  & $\epsilon_{cb}^R$  & $\epsilon_{\tau \nu}^L$  & $\epsilon_{\tau \nu}^R$    \\ \hline
\multirow{4}{*}{2P}
&($\epsilon_{cb}^L,\,\epsilon_{\tau \nu}^L$)                          & -0.045&  0.032  & 1.53  & 0.21  &1.46  & -0.93   & -0.27 &5.49  & -0.345 &    ---  & -0.276& ---   \\ \cline{2-13}
&($\epsilon_{cb}^L,\,\epsilon_{\tau \nu}^R$)                          & -0.047&  0.027  & 1.53  & -0.013 &1.46 & -0.94   & -0.27 &5.44  &  0.584 &    ---  & ---    & 0.897\\ \cline{2-13}
&($\epsilon_{cb}^R,\,\epsilon_{\tau \nu}^L$)                          & 2.57  &  0.46   & 1.55   & 0.19   &1.52 & -0.059 & -0.26 &12.28 &  ---  & -0.322 &  0.271& ---   \\ \cline{2-13}
&($\epsilon_{cb}^R,\, \epsilon_{\tau \nu}^R$)                         & -0.047&  0.027  & 1.53   & -0.013 &1.46 & -0.121  & -0.27 &5.44  &  ---  &  0.584 &   ---  & 0.897\\ \hline\hline
\multirow{2}{*}{3P}
&($\epsilon_{cb}^L,\,\epsilon_{cb}^R,\,\epsilon_{\tau \nu}^L$)        & 0.31  &  -0.20 & 1.52   & 0.22   &1.41 & -0.91 & -0.27  &5.34  & 0.272  & -0.051 & 0.326 & ---   \\ \cline{2-13}
&($\epsilon_{cb}^R,\,\epsilon_{\tau \nu}^L,\,\epsilon_{\tau \nu}^R$)  & 0.31  &  -0.21 & 1.52   & 0.011  &1.41 & -0.91 & -0.27 &5.29  & ---     & 0.466  &-0.038 & 1.082\\ \cline{2-13}
&($\epsilon_{cb}^L,\,\epsilon_{\tau \nu}^L,\,\epsilon_{\tau \nu}^R$)  & -0.048&  0.027 & 1.53   & $-7.4\times 10^{-7}$ &                      
                                                                                                               1.46 & -0.94 & -0.27 &5.44    & 0.666  & ---     & 0.008 & 0.764 \\ \hline\hline
4P
&($\epsilon_{cb}^L,\,\epsilon_{cb}^R,\,\epsilon_{\tau \nu}^L,\,\epsilon_{\tau \nu}^R$) 
                                                                      & 0.31  & -0.21  & 1.52   & $-4.1\times 10^{-6}$ & 
                                                                                                               1.41 & -0.91 & -0.27 &5.29  & 1.016  & -0.105 & 0.009 & -0.469\\ \hline
\end{tabular}          
\caption{By turning on the parameters of the second column~[keeping the remaining parameters of the Lagrangian~\eqref{BRBc_modified} equal to zero], we obtain several effective models at low energies. The models in rows 3-6 have two free parameters (the chiral couplings), and in the following, they will be referred to as 2P models. In the same sense, we will refer to the models in the rows 7-9 as 3P models. In the last row, the model with all the parameters turned on is shown . The pulls for each observable are shown in columns 3-8, the minimum value for the $\chi^2$ is shown in column 9,  and the best-fit point for the chiral charges is shown for each model in the last four columns, for  a gauge boson mass  $M_{W^\prime}= 1$ TeV. All  2P models have an acceptable value for  $ \chi^2_ {\text{min}}/\text{dof}\sim 1$, except the model with RH coupling to quarks and LH coupling to leptons. The goodness of fit decreases for the 3P and 4P models since, for them, the number of parameters increases while  $\chi^2_{\text{min}}$  stays at nearly the same value.}
\label{tab:pulls}
\end{center}
\end{table}


As the next step in our analysis, we will explore in a more detailed way the four 2P models $LL$, $LR$, $RL$, and $RR$, which according to our $\chi^2$ analysis are the best candidates to address the charged-current $B$ anomalies. By considering two different datasets, $R(D)$ and $R(D^\ast)$, and all of the $b \to c\tau \bar{\nu}_\tau$ observables, we determine the regions in the parameter space favored by the experimental data.

\subsection{$LL$ scenarios ($C^V_{LL}\neq 0$)}
In this scenario, we consider a $W^\prime$ boson that couples only to LH quark and LH lepton currents inducing the semi-tauonic operator $(\bar{c} \gamma_\mu P_L b) (\bar{\tau} \gamma^\mu P_L \nu_{\tau})$, i.e., $C^V_{LL}\neq 0$. In Figs.~\ref{ParameterSpace1}(a) and~\ref{ParameterSpace1}(b), we show the 95\% confidence level (C.L.) allowed parameter space in the ($\epsilon^L_{cb},\epsilon^L_{\tau\nu_\tau}$) plane, associated with the couplings in Eq.~\eqref{CLL}, for $M_{W^\prime}= 0.5$ TeV and $M_{W^\prime}= 1$ TeV, respectively. In order to see the impact of the polarization measurements~\cite{Hirose:2017dxl,Hirose:2016wfn,Abdesselam:2019wbt}, the purple region is obtained by considering the HFLAV-2019 averages on $R(D)$ and $R(D^\ast)$~\cite{HFLAVsummer}, while the green region is obtained by taking into account all the $b \to c\tau \bar{\nu}_\tau$ observables, namely $R(D^{(\ast)})$, $R(J/\psi)$, $F_L(D^*)$, $P_\tau(D^*)$ (see Table~\ref{Table1}), and considering the upper limit ${\rm BR}(B_c^- \to \tau^- \bar{\nu}_{\tau}) < 10\%$. It is observed that the allowed region for $R(D^{(\ast)})$ is significantly reduced to two symmetrical regions when all of the $b \to c\tau \bar{\nu}_\tau$ observables are considered. This is due mainly to the effect of the polarization $F_L(D^*)$, whereas observables such as $R(J/\psi)$ and $P_\tau(D^*)$ have little influence due to their large experimental uncertainties. This effect is in agreement with the analysis presented in Ref.~\cite{Shi:2019gxi}.
It is remarkable that the $R(D^{(\ast)})$ HFLAV-2019 averages allow the solution $(\epsilon^L_{\tau\nu_\tau},\epsilon^L_{cb})= (0,0)$; this result is consistent with the SM and does not require NP explanations.

\begin{figure*}[!t]
\centering
\includegraphics[scale=0.45]{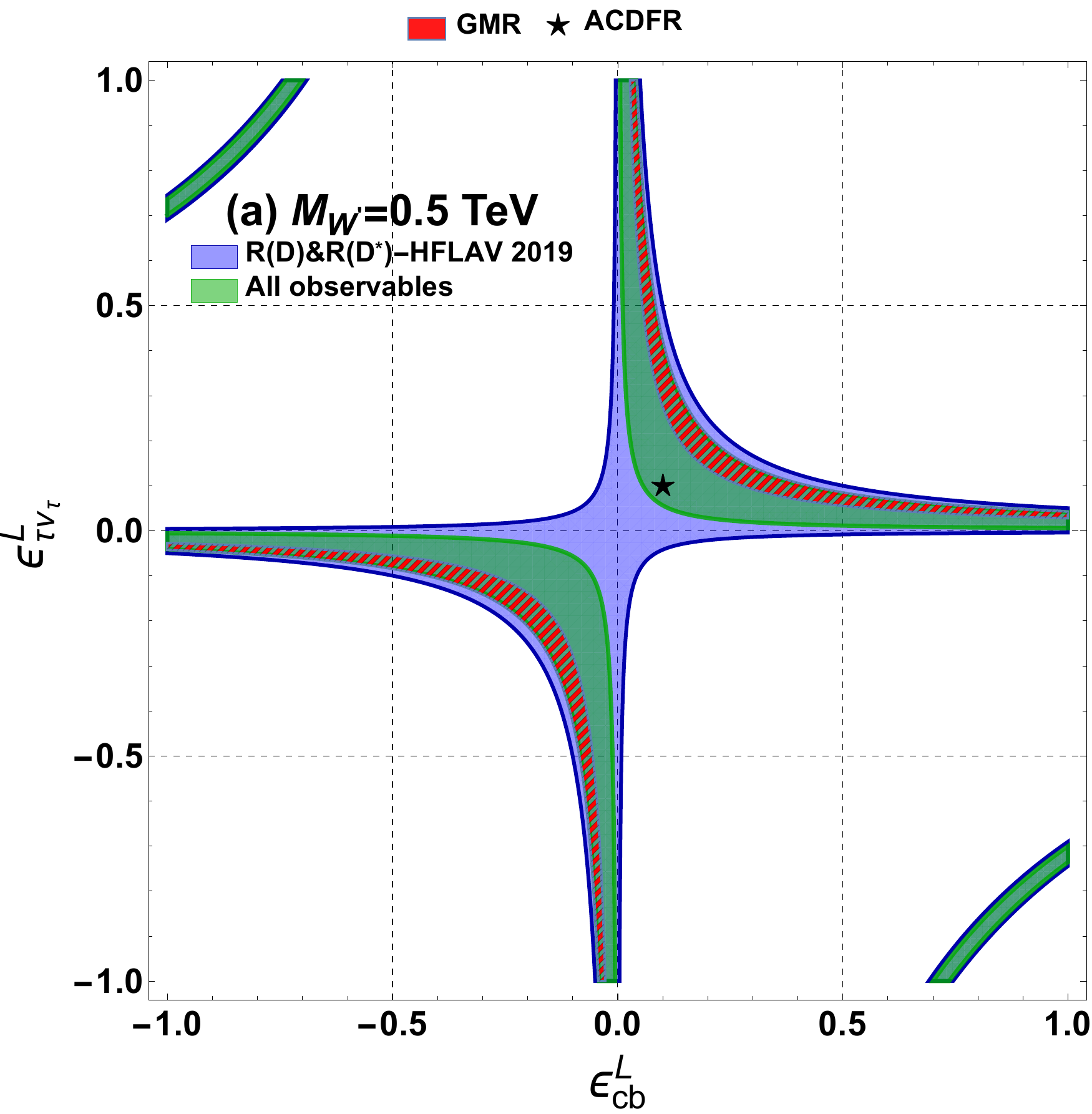} \
\includegraphics[scale=0.45]{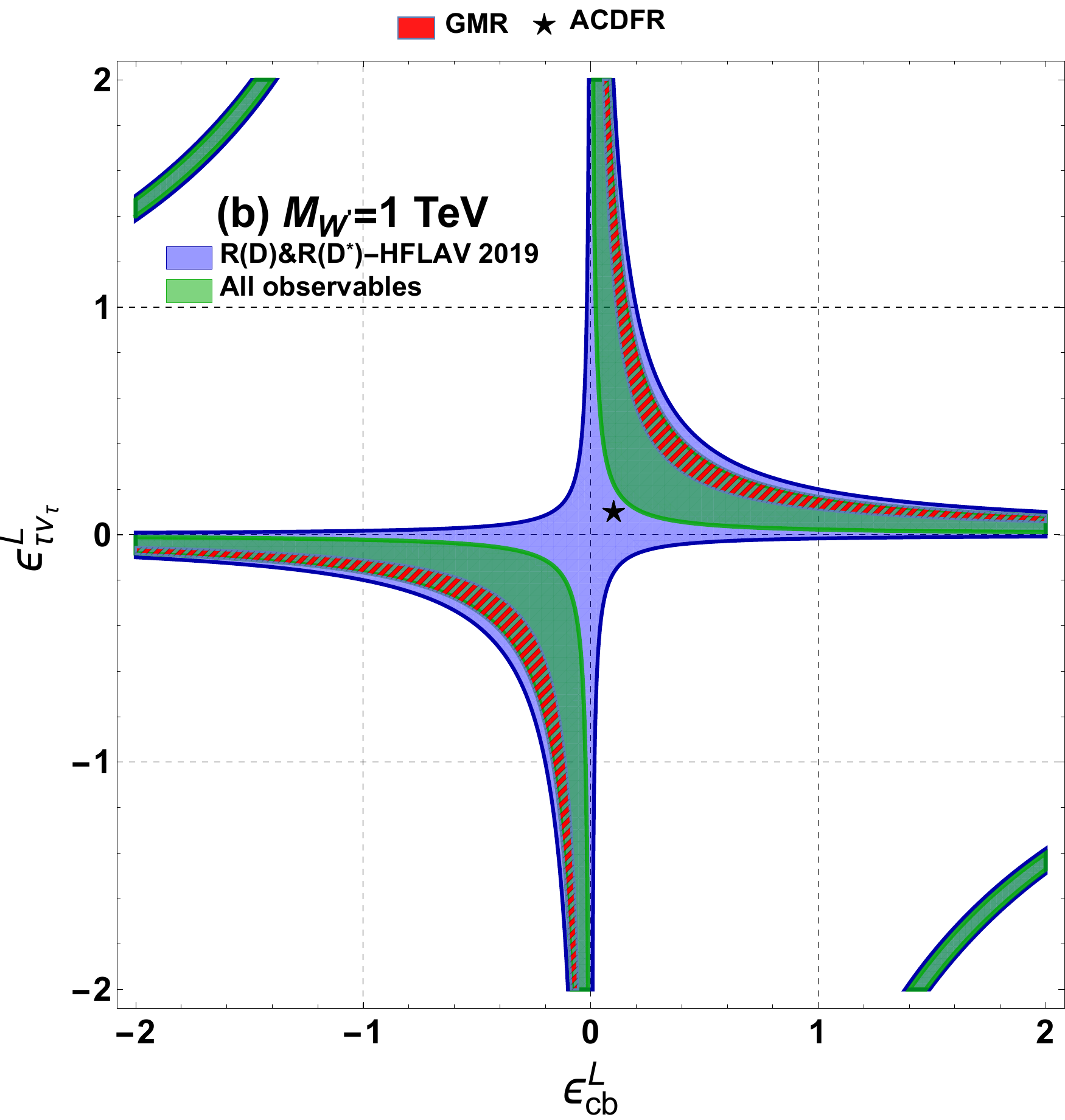}
\caption{\small The 95\% CL allowed parameter space in the ($\epsilon^L_{cb},\epsilon^L_{\tau \nu_\tau}$) plane for $(a) \ M_{W^\prime} = 0.5$ TeV and $(b) \ M_{W^\prime} = 1$ TeV. The purple region is obtained by considering only $R(D^{(\ast)})$ from HFLAV-2019 average, while the green one is obtained by taking into account all of the $b \to c\tau \bar{\nu}_\tau$ observables. The black star and red hatched region represent the ACDFR model~\cite{Abdullah:2018ets} and GMR analysis~\cite{Greljo:2018tzh}, respectively. See the text for details.}
\label{ParameterSpace1}
\end{figure*}

In order to improve our analysis, we will include some of the benchmark models that have already been studied in the literature~\cite{Greljo:2015mma,Faroughy:2016osc,Abdullah:2018ets,Greljo:2018tzh}: 
\begin{itemize}
\item In Ref.~\cite{Abdullah:2018ets}, Abdullah, Calle, Dutta, Flor\'{e}z, and Restrepo considered a simplified $W^\prime$ model (referred to by us as the ACDFR model)  which preferentially couples to the bottom and charm quarks and $\tau$ leptons, through the NP couplings $g_q^\prime$ and $g_\ell^\prime$, respectively. They showed that for $W^\prime$ masses in the range $[250, 750]$ GeV and  couplings $g_q^\prime = g_\ell^\prime = 0.1$, such scenario could be probed at the LHC with a luminosity of $100 \ \rm{fb}^{-1}$. This model is represented in Figs.~\ref{ParameterSpace1}(a) and~\ref{ParameterSpace1}(b) by the black star. We notice that for $M_{W^\prime} = 0.5$ TeV the ACDFR model is enabled both for HFLAV-2019 and all observables, while for $M_{W^\prime} = 1$~TeV, it is still allowed by HFLAV-2019.  
\item In Ref.~\cite{Greljo:2018tzh}, Greljo, Martin, and Ruiz performed a study (referred to by us as the GMR analysis) of the connection between NP scenarios addressing the $R(D^{(\ast)})$ anomalies and the mono-tau signature at the LHC, $pp \to \tau_h X + \rm{MET}$. By using current ATLAS~\cite{Aaboud:2018vgh} and CMS~\cite{Sirunyan:2018lbg} data they constrained different scenarios --particularly those regarding a $W^\prime$ boson scenario-- and they found that~\cite{Greljo:2018tzh}
\begin{equation}
\epsilon_{cb}^{L} \epsilon_{\tau \nu_\tau}^{L} = (0.14 \pm 0.03) \bigg(\frac{M_{W^\prime}}{{\rm TeV}} \bigg)^{2} ,  
\end{equation}

\noindent for $W^\prime$ masses in the range $[0.5, 3.5]$ TeV, which is in agreement with the value $\epsilon_{cb}^{L} \epsilon_{\tau \nu_\tau}^{L} = 0.107 \ (M_{W^\prime}/ \rm{TeV})^2$ obtained in~\cite{Iguro:2018fni}. This result is represented by the red hatched region in Figs.~\ref{ParameterSpace1}(a) and~\ref{ParameterSpace1}(b). We can appreciate that the allowed parameter region by $R(D^{(\ast)})$ HFLAV 2019 and all $b \to c\tau \bar{\nu}_\tau$ observables of our analysis are consistent and overlap with this region.
\item In Refs.~\cite{Greljo:2015mma,Faroughy:2016osc} the authors introduced a color-neutral $SU(2)_L$ triplet of massive vector bosons that couple predominantly with third generation fermions (both quarks $g_q$ and $g_\ell$ leptons), with an underlying dynamics generated by an approximated $U(2)_q \times U(2)_\ell$ flavor symmetry; however, in light of new experimental measurements, this model is disfavored unless a fine-tuning of the couplings is carried out.
\end{itemize}





\subsection{$RR$ scenarios ($C^V_{RR}\neq 0$)}

\begin{figure*}[!t]
\centering
\includegraphics[scale=0.4]{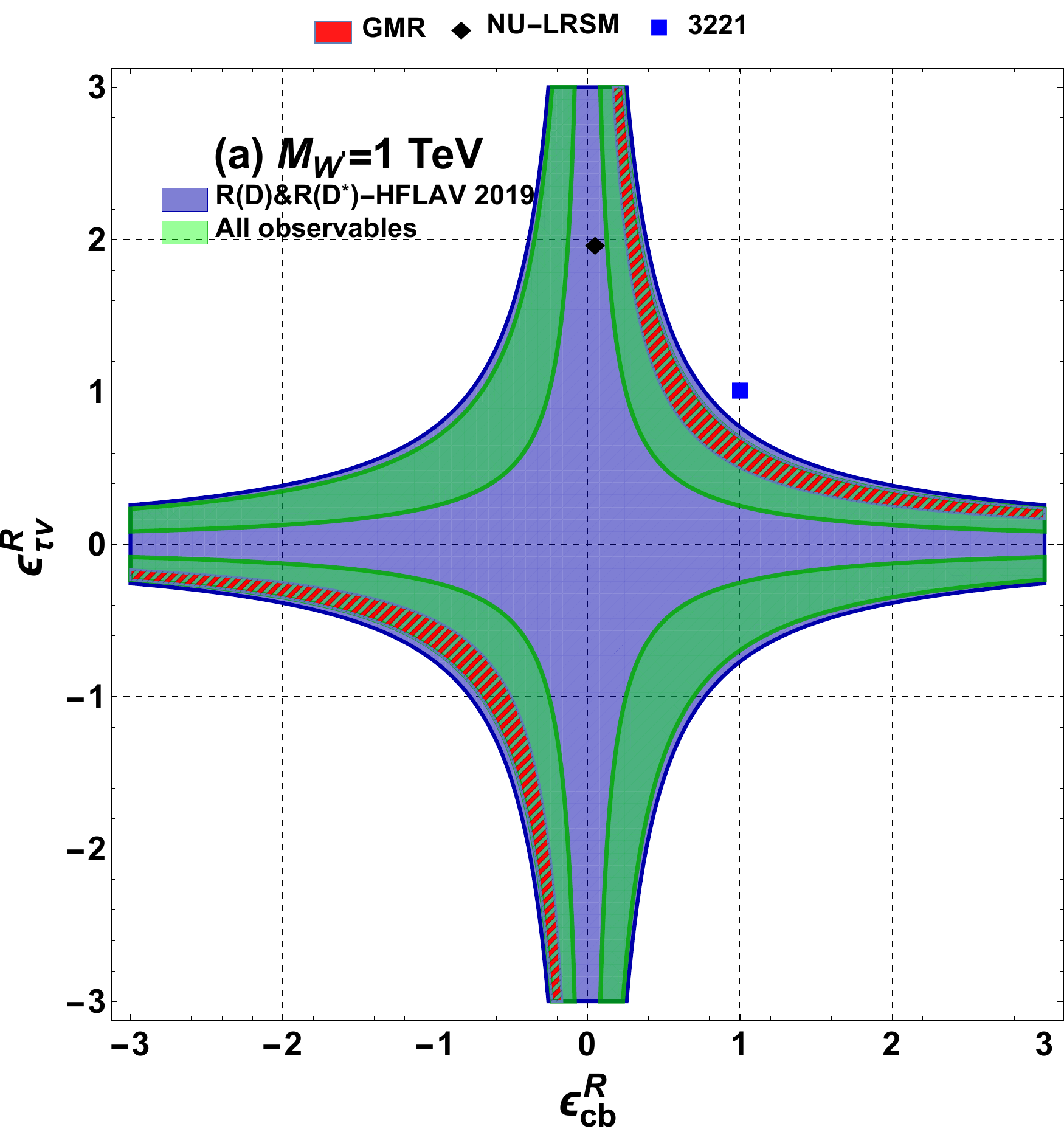}  \
\includegraphics[scale=0.4]{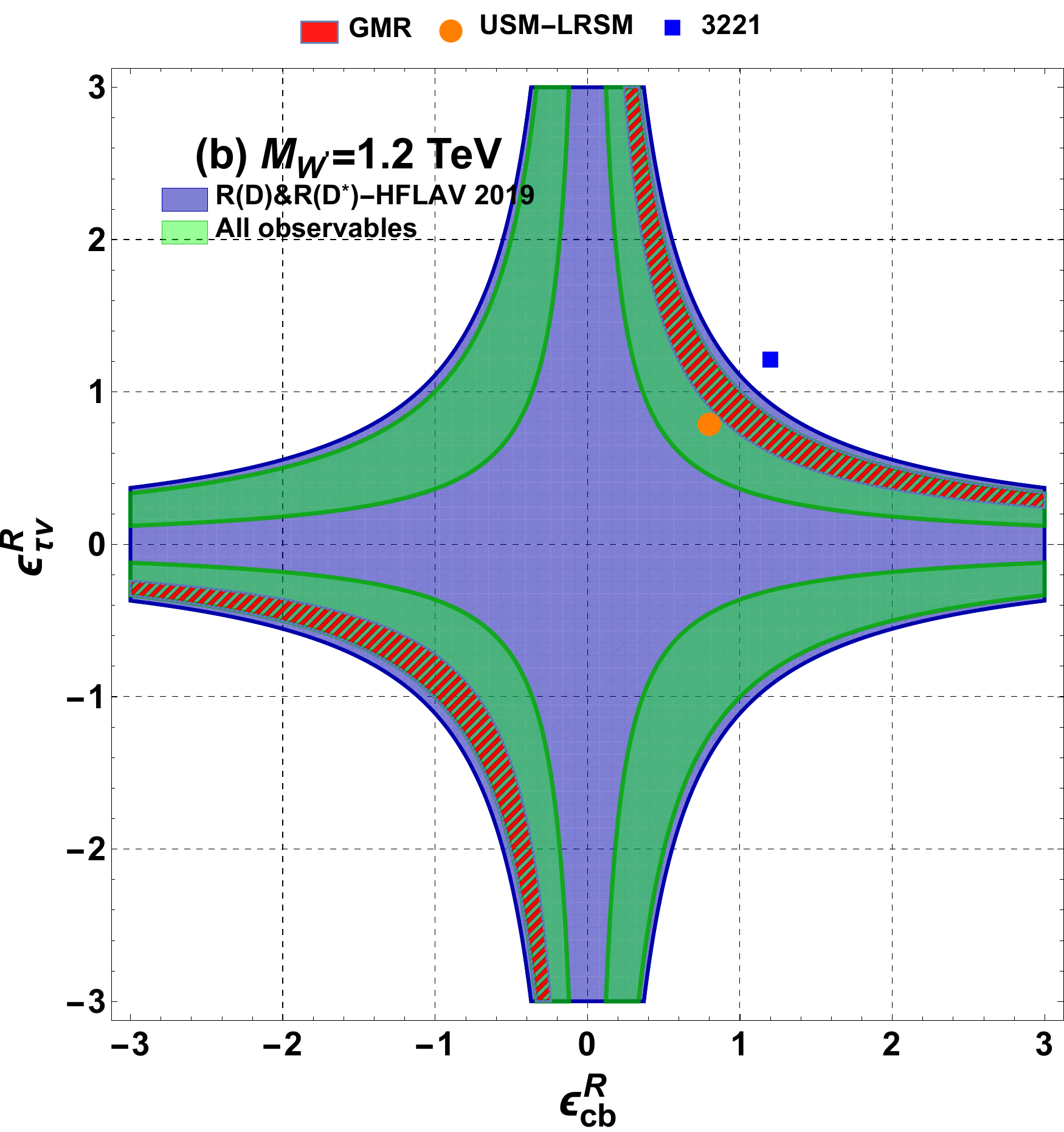} \
\caption{\small The 95\% C.L. allowed parameter space in the ($\epsilon^R_{cb},\epsilon^R_{\tau \nu_\tau}$) plane for  $(a) \ M_{W^\prime} = 1$ TeV and  $(b) \ M_{W^\prime} = 1.2$ TeV. The purple region is obtained by considering only $R(D^{(\ast)})$ from HFLAV-2019 averages, while the green one is obtained by taking into account all of the $b \to c\tau \bar{\nu}_\tau$ observables. The black diamond, blue squared, red hatched region, and orange circle represent the NU-LRSM model~\cite{He:2012zp,He:2017bft}, 3321 gauge model~\cite{Asadi:2018wea,Greljo:2018ogz}, GMR analysis~\cite{Greljo:2018tzh}, and USM-LRSM~\cite{Babu:2018vrl}, respectively. See the text for details.}
\label{ParameterSpace2}
\end{figure*}

In this scenario, the $W^\prime$ is the gauge boson associated with the interaction between the RH quark and RH lepton currents involving a RH sterile neutrino. This RH current interpretation to the $R(D^{(\ast)})$ anomalies have been discussed recently in the literature within different NP realizations~\cite{Li:2018rax,Carena:2018cow,Babu:2018vrl,Asadi:2018wea,Greljo:2018ogz,Robinson:2018gza,Asadi:2018sym,Azatov:2018kzb,Cvetic:2017gkt}. We plot in Figs.~\ref{ParameterSpace2}(a) and~\ref{ParameterSpace2}(b) the 95\% C.L. allowed parameter space in the ($\epsilon^R_{cb},\epsilon^R_{\tau\nu_\tau}$) plane for masses $M_{W^\prime}= 1$ TeV and 1.2 TeV, respectively. The purple and green regions are obtained by taking into account only $R(D)$ and $R(D^\ast)$ from HFLAV-2019 averages~\cite{HFLAVsummer} and all the $b \to c\tau \bar{\nu}_\tau$ observables, respectively. It is found that the allowed region for $R(D^{(\ast)})$ is significantly reduced to four-fold symmetrical regions when all of the $b \to c\tau \bar{\nu}_\tau$ observables are considered. As in the $LL$ scenarios previously discussed, this is mainly due to the effect of the polarization $F_L(D^*)$. For further discussion, we consider some benchmark models:

\begin{itemize}
\item The authors of Refs.~\cite{Asadi:2018wea,Greljo:2018ogz} presented a model where the SM is extended by the gauge group $SU(3)_C \times SU(2)_L \times SU(2)_V \times U(1)^\prime$, with $g_V$ and $g^\prime$ being the corresponding new gauge couplings. After the spontaneous symmetry breaking $SU(2)_V \times U(1)^\prime \to U(1)_Y$, new heavy vector bosons are generated. In addition, the SM fermion content is accompanied by new heavy vector-like fermions (both quarks and leptons) that mix with the RH fermions of the SM, which is required in order to provide an explanation of the $R(D^{(\ast)})$ anomalies. Since the results in~\cite{Asadi:2018wea,Greljo:2018ogz} are very similar, for simplicity, we will consider the analysis of Ref.~\cite{Greljo:2018ogz} for comparison (referred to as the 3221 gauge model). Translating the notation in~\cite{Greljo:2018ogz} into ours, we have $\epsilon_{cb}^{R} = g_V c_q^{23}$ and $\epsilon_{\tau \nu_\tau}^{R} = g_V c_N^3$, with $c_q^{23}, c_N^{3}$ coefficients that encode the flavor dependence. Given that $M_{W^\prime} = g_V v_V /2$~\cite{Greljo:2018ogz}, a viable $1\sigma$ solution to the anomalies is obtained for a vacuum expectation value (VEV) of $v_V \simeq 2000$ GeV, $g_V \simeq \mathcal{O}(1-3)$ and $c_q^{23}= c_N^{3} \simeq 1$, implying $W^\prime$ masses in the range $1000 \lesssim M_{W^\prime} ({\rm GeV}) \lesssim 3000$ to avoid the perturbative limit~\cite{Greljo:2018ogz}. By taking representative values of $v_V \simeq 2000$ GeV and $g_V \simeq 1 - 1.2$, the 3221 gauge model is depicted by the blue squared in Figs.~\ref{ParameterSpace2}(a) and~\ref{ParameterSpace2}(b) for $M_{W^\prime} = 1$ TeV and  $1.2$ TeV, respectively. According to our analysis, this model is disfavored by the new data. We have also checked, that for $W^\prime$ masses higher than 1.2~TeV, this is still disfavored. However, as discussed in~\cite{Robinson:2018gza}, there is a freedom in the flavor structure of the $c_q^{23}, c_N^{3}$ couplings, and it is possible to get, in general, different values $c_q^{23} \neq c_N^{3}$ than the ones assumed in~\cite{Greljo:2018ogz}.
\item In the GMR analysis~\cite{Greljo:2018tzh} previously discussed, the authors also found that for RH $W^\prime$ models the solution is
\begin{equation}
\epsilon_{cb}^{R} \epsilon_{\tau \nu_\tau}^{R} = (0.6 \pm 0.1) \ \bigg(\frac{M_{W^\prime}}{{\rm TeV}} \bigg)^{2},  
\end{equation}
\noindent which is represented by the red hatched region in Figs.~\ref{ParameterSpace2}(a) and~\ref{ParameterSpace2}(b), which is consistent with the value $\epsilon_{cb}^{R} \epsilon_{\tau \nu_\tau}^{R} = 0.55 \ (M_{W^\prime}/ \rm{TeV})^2$ obtained in~\cite{Iguro:2018fni}. Again, the allowed parameter region by $R(D^{(\ast)})$ HFLAV 2019 and all $b \to c\tau \bar{\nu}_\tau$ observables of our analysis are consistent and overlap with this region.
\item In Refs.~\cite{He:2012zp,He:2017bft}, the anomalies have been addressed within the framework of the non-universal left-right symmetric model (NU-LRSM) with enhanced couplings to the third generation. In terms of our notation, we have the result that in the NU-LRSM, the effective couplings are $\epsilon_{cb}^{R} = g_R |V_{Rcb}|$ and $\epsilon_{\tau \nu_\tau}^{R} = g_R |V^{\ell}_{R3\tau}|$, with $g_R$ being the RH gauge coupling, $V_{Rcb}$ and $V^{\ell}_{R3\tau}$ the RH quark and lepton mixing element, respectively. It is assumed that, taking $M_{W^\prime} \simeq 1$~TeV, $g_R \simeq 1$, $|V^{\ell}_{R3\tau}| \simeq 1$, and $|V_{Rcb}| \simeq |V_{cb}|$~\cite{He:2012zp,He:2017bft}, as shown by the black diamond in Fig.~\ref{ParameterSpace2}(a), the model accommodates the tension in $R(D^{(\ast)})$. One can observe that this framework is still allowed by the HFLAV-2019 average, but not with all observables data.
\item A class of LRSM (parity symmetric and asymmetric) that implemented vector-like fermions to generate quark and lepton masses via a universal seesaw mechanism (USM) have been studied in Ref.~\cite{Babu:2018vrl} to explain the anomalies. In the USM-LRSM, the mass of the RH charged gauge boson is given by $M_{W^\prime} = M_{W_R}= g_R \kappa_R /\sqrt{2}$, with $\kappa_R \sim 2$~TeV being the VEV of the neutral member of the doublet $\chi_R$ (for details, see Ref.~\cite{Babu:2018vrl}), and the effective couplings are simply $\epsilon_{cb}^{R} = g_R /\sqrt{2}$ and $\epsilon_{\tau \nu_\tau}^{R} = g_R/\sqrt{2}$. Taking the lower mass limit $M_{W_R} \simeq 1.2$ TeV (obtained for the parity asymmetric case~\cite{Babu:2018vrl}), the USM-LRSM is represented by the orange circle in Fig.~\ref{ParameterSpace2}(b). This setup is allowed by both $R(D^{(\ast)})$ and all of the $b \to c\tau \bar{\nu}_\tau$ observables. 
\end{itemize}


\FloatBarrier

\subsection{$RL$ and $LR$ scenarios ($C^V_{RL}\neq 0$ and $C^V_{LR}\neq 0$)}
Finally, we consider a class of scenarios where the quark and lepton currents with different quiralities projection couple to the $W^\prime$ boson, i.e., semi-tauonic operators of the types $(\bar{c} \gamma_\mu P_R b) (\bar{\tau} \gamma^\mu P_L \nu_{\tau})$ and $(\bar{c} \gamma_\mu P_L b) (\bar{\tau} \gamma^\mu P_R \nu_{\tau})$ that implies $C^V_{RL}\neq 0$ and $C^V_{LR}\neq 0$, respectively. For a representative mass value of $M_{W^\prime} = 1$ TeV, we display in Fig.~\ref{ParameterSpace3} the 95\% C.L. allowed parameter space for the couplings in the $LR$ (left panel) and $RL$ (right panel) scenarios. The case of $M_{W^\prime} \geq 1$~TeV requires higher effective coupling values. For the $LR$ case, it can be inferred that the allowed region for $R(D^{(\ast)})$ is reduced to four-fold symmetrical regions when all of the $b \to c\tau \bar{\nu}_\tau$ observables are considered. While for the $RL$ case, the permitted region is barely reduced when all observables are taken into account. In both scenarios it is found that a NP solution $(0,0)$ is admissible.

So far, particular NP models realization of such $LR$ and $RL$ scenarios have not been studied in the literature. However, interestingly enough, recently Bhattacharya \textit{et al}~\cite{Bhattacharya:2019olg} have explored the possibility of how the measurement of CP-violating observables in $\bar{B}^0 \to D^{\ast +} \mu^- \bar{\nu}_\mu$ can be used to differentiate the NP scenarios. Particularly, they found that the only way to generate sizable $CP$-violating effects is with LH and RH $W^\prime$ bosons (with sizable mixing) that contribute to $b \to c\ell \bar{\nu}_\ell$~\cite{Bhattacharya:2019olg}.

\begin{figure*}[!t]
\centering
\includegraphics[scale=0.4]{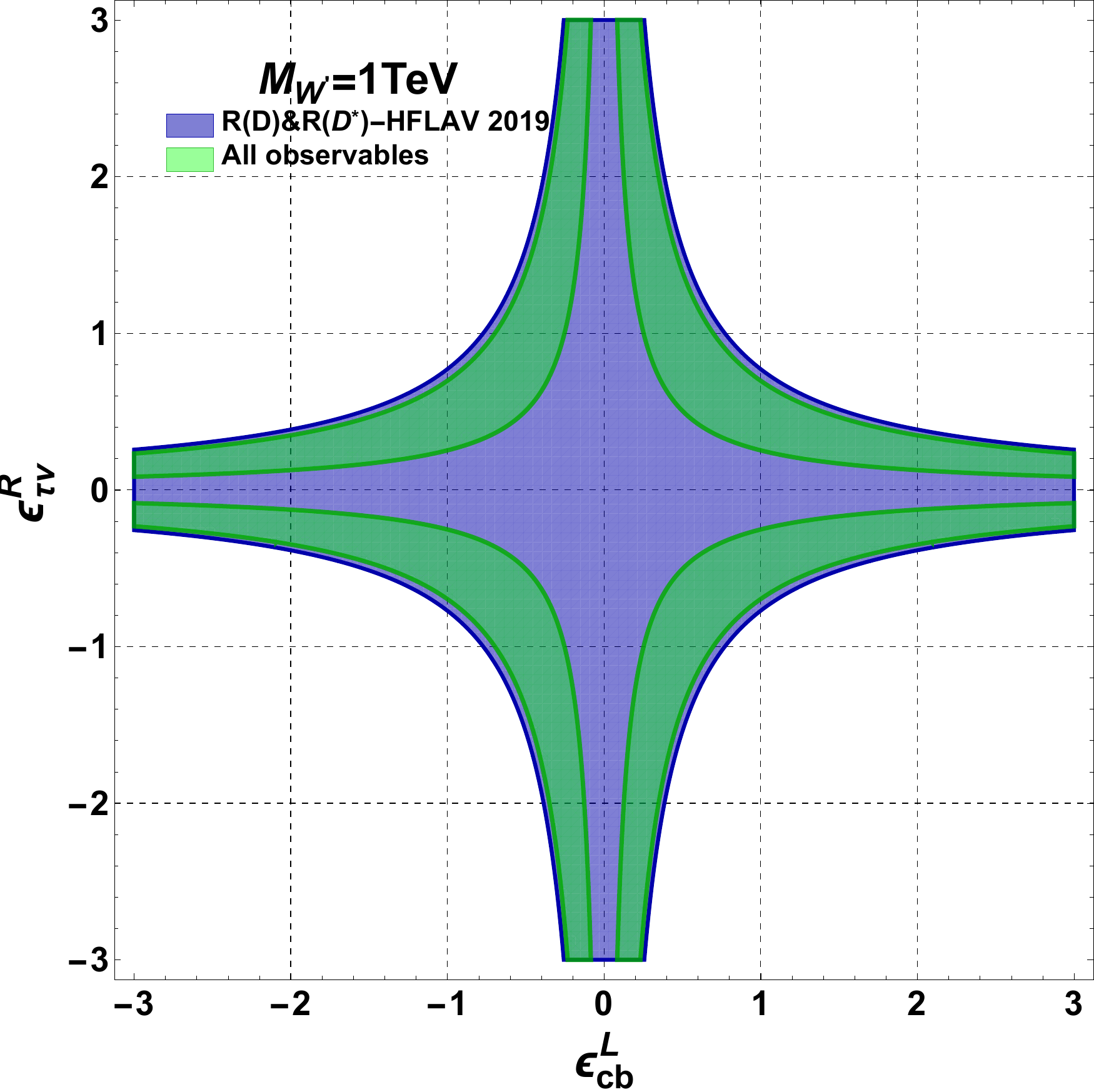}  \
\includegraphics[scale=0.4]{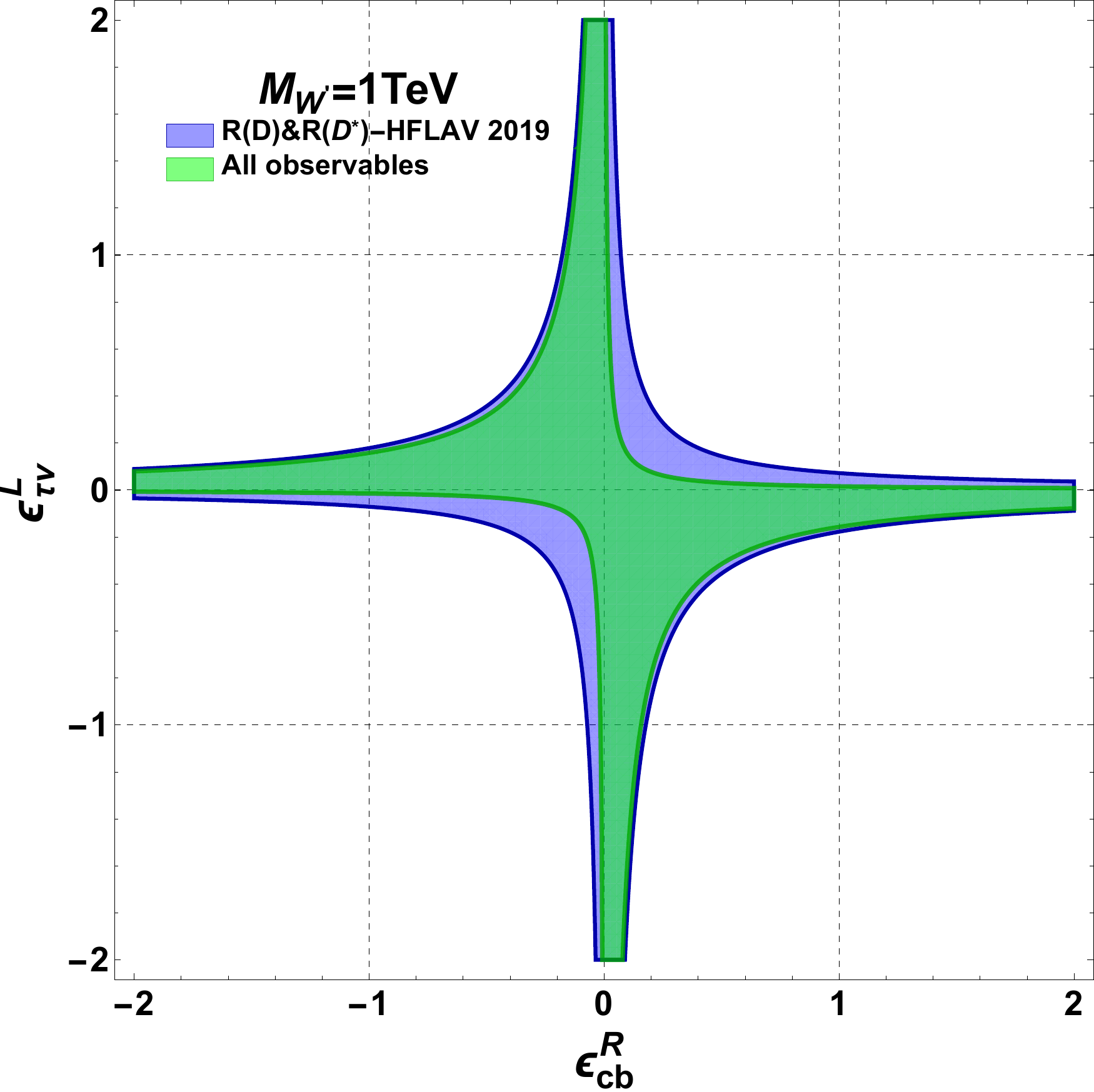} 
\caption{\small The 95\% C.L. allowed parameter space in the ($\epsilon^L_{cb},\epsilon^R_{\tau \nu_\tau}$) and ($\epsilon^R_{cb},\epsilon^L_{\tau \nu_\tau}$) planes for a mass value of $M_{W^\prime} = 1$ TeV.}
\label{ParameterSpace3}
\end{figure*}

\section{Conclusions} \label{Conclusion}

Motivated by the new HFLAV world average values on the ratios $R(D^{(\ast)})$, due to the recent Belle measurements, we addressed the anomalies $R(D^{(*)})$ related to the charged-current transition $b \to c \tau \bar{\nu}_\tau$ within a general $W'$ boson scenario. In order to provide a robust analysis, we considered in addition the available experimental information on all of the charged transition $b \to c \tau \bar{\nu}_\tau$ observables, namely the ratios $R(J/\psi)$, $R(X_c)$, and the polarizations $P_\tau(D^\ast), F_L(D^\ast)$, as well as the upper limit BR$(B_c^{-} \to \tau^{-} \bar{\nu}_\tau) < 10 \%$.
We carried out a model-independent study, based on the most general effective Lagrangian given in terms of the flavor-dependent couplings $\epsilon_{cb}^{L,R}$ and $\epsilon_{\tau\nu_\tau}^{L,R}$ of the currents $\bar{c} \gamma_\mu P_{L,R} b$ and $\bar{\tau} \gamma^\mu P_{L,R} \nu_{\tau}$, that yields to a tree-level effective contribution generated by a general $W'$ boson. With the above-mentioned observables, we performed a $\chi^2$ analysis by considering the cases of two, three, and four nonzero $\epsilon_{cb}^{L,R}$ and $\epsilon_{\tau\nu_\tau}^{L,R}$ couplings (with different chiral charges), referred to as the 2P, 3P, and 4P models, respectively. It is found that the 2P models represent the best candidate to adjust the experimental charged current $B$ anomalies. 



Next, we studied all of the possible combinations of 2P models ($LL$, $RR$, $LR$, and $RL$ scenarios) and took into account two different datasets: $R(D^{(\ast)})$ only and all $b \to c\tau \bar{\nu}_\tau$ observables; we determined the regions in parameter space favored by these observables for different values of the $W'$ boson mass preferred in the literature. For the $LL$ and $RR$ scenarios, we obtained that part of the allowed parametric space is consistent with the mono-tau signature $pp \to \tau_h X + \rm{MET}$ at the LHC. In order to improve the discussion, we included in our analysis some of the $W'$ boson NP realizations that have already been studied in the $LL$ and $RR$ scenarios. We found which of these benchmark models are favored or disfavored by the new data. Regarding the $LR$ and $RL$ scenarios, as far as we know, these have not been previously reported in the literature, and our results showed that it would be interesting to study a particular NP model since this could generate $CP$-violating effects in $\bar{B}^0 \to D^{\ast +} \mu^- \bar{\nu}_\mu$, as discussed in~\cite{Bhattacharya:2019olg}.

\acknowledgments
We are grateful to William A. Ponce and Carlos E. Vera for their contributions at the early stage of this work.  
N. Q. acknowledges support from the Direcci\'{o}n General de Investigaciones, Universidad Santiago de Cali, under Project No. 935-621118-3.



\begin{thebibliography}{99}
\bibitem{Ciezarek:2017yzh}
G.~Ciezarek, M.~Franco Sevilla, B.~Hamilton, R.~Kowalewski, T.~Kuhr, V.~L\"{u}th and Y.~Sato, A Challenge to Lepton Universality in B Meson Decays, Nature {\bf 546}, 227 (2017). \href{http://arxiv.org/abs/1703.01766}{[arXiv:1703.01766 [hep-ex]]}

\bibitem{Bifani:2018zmi} 
 S.~Bifani, S.~Descotes-Genon, A.~Romero Vidal and M.~H.~Schune, Review of Lepton Universality tests in $B$ decays,  J.\ Phys.\ G {\bf 46}, 023001 (2019). \href{http://arxiv.org/abs/1809.06229}{[arXiv:1809.06229 [hep-ex]]}

\bibitem{Lees:2012xj}
B. Aubert \textit{et al.} (BABAR Collaboration), Evidence for an excess of $B \to D^{(*)} \tau \nu$ decays, 
Phys. Rev. Lett. \textbf{109}, 101802 (2012). \href{http://arxiv.org/abs/1205.5442}{[arXiv:1205.5442 [hep-ex]]}

\bibitem{Lees:2013uzd}
J. Lees \textit{et al.} (BABAR Collaboration), Measurement of an Excess of $B \to D^{(*)} \tau \nu$ Decays and Implications for Charged Higgs Bosons, Phys. Rev. D \textbf{88}, 072012 (2013). 
\href{http://arxiv.org/abs/1303.0571}{[arXiv:1303.0571 [hep-ex]]}



\bibitem{Fajfer:2012vx} 
S. Fajfer, J. F. Kamenik, and I. Nisandzi\'c, On the $B \to D^*\tau \bar{\nu}_\tau$ sensitivity to new physics, 
 Phys. Rev. D \textbf{85}, 094025 (2012). \href{http://arxiv.org/abs/1203.2654}{[arXiv:1203.2654 [hep-ph]]}
 
\bibitem{Fajfer:2012jt} 
S. Fajfer, J. F. Kamenik, I. Nisandzi\'c, and J. Zupan, Implications of lepton flavor universality violations in B decays, Phys. Rev. Lett. \textbf{109}, 161801 (2012). \href{http://arxiv.org/abs/1206.1872}{[arXiv:1206.1872 [hep-ph]]}

\bibitem{Bailey:2012jg} 
J. A. Bailey \textit{et al.}, Refining new-physics searches in $B \to D\tau\nu$ decay with lattice QCD,
Phys. Rev. Lett. \textbf{109}, 071802 (2012). \href{http://arxiv.org/abs/1206.4992}{[arXiv:1206.4992 [hep-ph]]}

\bibitem{Huschle:2015rga} 
M.~Huschle {\it et al.} (Belle Collaboration), Measurement of the branching ratio of $\bar{B} \to D^{(\ast)} \tau^- \bar{\nu}_\tau$ relative to $\bar{B} \to D^{(\ast)} \ell^- \bar{\nu}_\ell$ decays with hadronic tagging at Belle, Phys.\ Rev.\ D {\bf 92}, 072014 (2015). \href{http://arxiv.org/abs/1507.03233}{[arXiv:1507.03233 [hep-ex]]}
  
\bibitem{Sato:2016svk} 
  Y.~Sato {\it et al.} (Belle Collaboration), Measurement of the branching ratio of $\bar{B}^0 \rightarrow D^{*+} \tau^- \bar{\nu}_{\tau}$ relative to $\bar{B}^0 \rightarrow D^{*+} \ell^- \bar{\nu}_{\ell}$ decays with a semileptonic tagging method, Phys.\ Rev.\ D {\bf 94}, 072007 (2016). \href{http://arxiv.org/abs/1607.07923}{[arXiv:1607.07923 [hep-ex]]}

\bibitem{Hirose:2017dxl} 
S.~Hirose {\it et al.} (Belle Collaboration), Measurement of the $\tau$ lepton polarization and $R(D^*)$ in the decay $\bar{B} \rightarrow D^* \tau^- \bar{\nu}_\tau$ with one-prong hadronic $\tau$ decays at Belle, Phys.Rev. D \textbf{97}, 012004 (2018). \href{http://arxiv.org/abs/1709.00129}{[arXiv:1709.00129 [hep-ex]]}

\bibitem{Hirose:2016wfn} 
S.~Hirose {\it et al.} (Belle Collaboration),
Measurement of the $\tau$ lepton polarization and $R(D^*)$ in the decay $\bar{B} \to D^* \tau^- \bar{\nu}_\tau$, Phys.\ Rev.\ Lett.\  {\bf 118}, 211801 (2017) \href{http://arxiv.org/abs/1612.00529}{[arXiv:1612.00529 [hep-ex]]}.

\bibitem{Aaij:2015yra} 
R.~Aaij {\it et al.} (LHCb Collaboration), Measurement of the ratio of branching fractions $\mathcal{B}(\bar{B}^0 \to D^{*+}\tau^{-}\bar{\nu}_{\tau})/\mathcal{B}(\bar{B}^0 \to D^{*+}\mu^{-}\bar{\nu}_{\mu})$, Phys.\ Rev.\ Lett.\  {\bf 115}, 111803 (2015)  Erratum: [Phys.\ Rev.\ Lett.\  {\bf 115}, 159901 (2015)]
\href{http://arxiv.org/abs/1506.08614}{[arXiv:1506.08614 [hep-ex]]}

\bibitem{Aaij:2017deq} 
  R.~Aaij {\it et al.} (LHCb Collaboration), Test of Lepton Flavor Universality by the measurement of the $B^0 \to D^{*-} \tau^+ \nu_{\tau}$ branching fraction using three-prong $\tau$ decays,
Phys.\ Rev.\ D {\bf 97}, 072013 (2018) \href{http://arxiv.org/abs/1711.02505}{[arXiv:1711.02505 [hep-ex]]}.

\bibitem{Aaij:2017uff} 
R.~Aaij {\it et al.} (LHCb Collaboration), Measurement of the ratio of the $B^0 \to D^{*-} \tau^+ \nu_{\tau}$ and $B^0 \to D^{*-} \mu^+ \nu_{\mu}$ branching fractions using three-prong $\tau$-lepton decays,  Phys.\ Rev.\ Lett.\  {\bf 120}, 171802 (2018) \href{http://arxiv.org/abs/1708.08856}{[arXiv:1708.08856 [hep-ex]]}.

\bibitem{Amhis:2016xyh} 
Y.~Amhis {\it et al.}, Heavy Flavor Averaging Group (HFLAV), Averages of $b$-hadron, $c$-hadron, and $\tau$-lepton properties as of summer 2016,  	Eur. Phys. J. C \textbf{77}, 895 (2017). \href{http://arxiv.org/abs/1612.07233}{[arXiv:1612.07233 [hep-ex]]}.

\bibitem{HFLAVsummer}
For updated results, see the HFLAV average of $R(D^{(\ast)})$ for Spring 2019 at \url{https://hflav-eos.web.cern.ch/hflav-eos/semi/spring19/html/RDsDsstar/RDRDs.html}.

\bibitem{Bigi:2016mdz} 
D.~Bigi and P.~Gambino, Revisiting $B\to D \ell \nu$, Phys.\ Rev.\ D {\bf 94}, 094008 (2016).
\href{http://arxiv.org/abs/1606.08030}{[arXiv:1606.08030 [hep-ph]]}

\bibitem{Aoki:2016frl}
S. Aoki \textit{et al.}, (FLAG Working Group),  Review of lattice results concerning low-energy particle physics, Eur. Phys. J. C \textbf{77}, 112, (2017). \href{http://arxiv.org/abs/1607.00299}{arXiv:1607.00299 [hep-lat]}

\bibitem{Bernlochner:2017jka} 
F.~U.~Bernlochner, Z.~Ligeti, M.~Papucci, and D.~J.~Robinson, Combined analysis of semileptonic $B$ decays to $D$ and $D^*$: $R(D^{(*)})$, $|V_{cb}|$, and new physics, Phys.\ Rev.\ D {\bf 95}, 115008 (2017). \href{http://arxiv.org/abs/1703.05330}{arXiv:1703.05330 [hep-ph]}

\bibitem{Jaiswal:2017rve} 
S.~Jaiswal, S.~Nandi, and S.~K.~Patra, Extraction of $|V_{cb}|$ from $B\to D^{(*)}\ell\nu_\ell$ and the Standard Model predictions of $R(D^{(*)})$, JHEP {\bf 1712}, 060 (2017) \href{http://arxiv.org/abs/1707.09977}{[arXiv:1707.09977 [hep-ph]]}.
 
\bibitem{Bigi:2017jbd}
D.~Bigi, P.~Gambino ,and S.~Schacht, $R(D^*)$, $|V_{cb}|$, and the Heavy Quark Symmetry relations between form factors,   JHEP {\bf 11}, 061 (2017). \href{http://arxiv.org/abs/1707.09509}{[arXiv:1707.09509 [hep-ph]]}

\bibitem{Abdesselam:2019dgh} 
  A.~Abdesselam {\it et al.} (Belle Collaboration), Measurement of $\mathcal{R}(D)$ and $\mathcal{R}(D^{\ast})$ with a semileptonic tagging method, arXiv:1904.08794 [hep-ex].
  

\bibitem{Aaij:2017tyk} 
R.~Aaij {\it et al.} (LHCb Collaboration), Measurement of the ratio of branching fractions $\mathcal{B}(B_c^+\,\to\,J/\psi\tau^+\nu_\tau)$/$\mathcal{B}(B_c^+\,\to\,J/\psi\mu^+\nu_\mu)$,  Phys. Rev. Lett. \textbf{120}, 121801 (2018) \href{http://arxiv.org/abs/1711.05623}{[arXiv:1711.05623 [hep-ex]]}.

\bibitem{Dutta:2017xmj} 
R.~Dutta and A.~Bhol, $B_c \to (J/\psi,\,\eta_c)\tau\nu$ semileptonic decays within the standard model and beyond, Phys.\ Rev.\ D {\bf 96}, 076001 (2017). \href{http://arxiv.org/abs/1701.08598}{[arXiv:1701.08598 [hep-ph]]}

\bibitem{Watanabe:2017mip}  
R.~Watanabe,  New Physics effect on $B_c \to J/\psi \tau\bar\nu$ in relation to the $R_{D^{(*)}}$ anomaly, Phys.\ Lett.\ B {\bf 776}, 5 (2018). \href{http://arxiv.org/abs/1709.08644}{[arXiv:1709.08644 [hep-ph]]}

\bibitem{Murphy:2018sqg} 
C.~W.~Murphy and A.~Soni, Model-Independent Determination of $B_c^+ \to \eta_c\, \ell^+\, \nu$ Form Factors, Phys.\ Rev.\ D {\bf 98}, 094026 (2018)
\href{http://arxiv.org/abs/1808.05932}{[arXiv:1808.05932 [hep-ph]]}.

\bibitem{Cohen:2018dgz} 
T.~D.~Cohen, H.~Lamm and R.~F.~Lebed, Model-independent bounds on $R(J/\psi)$, JHEP {\bf 1809}, 168 (2018)
\href{http://arxiv.org/abs/1807.02730}{[arXiv:1807.02730 [hep-ph]]}.

\bibitem{Issadykov:2018myx} 
  A.~Issadykov and M.~A.~Ivanov, The decays $B_{c}\to J/\psi+\bar\ell\nu_\ell$ and $B_{c}\to J/\psi + \pi(K)$ in covariant confined quark model, Phys.\ Lett.\ B {\bf 783}, 178 (2018) [arXiv:1804.00472 [hep-ph]].

\bibitem{Azizi:2019aaf} 
  K.~Azizi, Y.~Sarac and H.~Sundu, Lepton flavor universality violation in semileptonic tree level weak transitions, Phys.\ Rev.\ D {\bf 99}, 113004 (2019) [arXiv:1904.08267 [hep-ph]].

  
\bibitem{Abdesselam:2019wbt} 
A.~Abdesselam {\it et al.} [Belle Collaboration], Measurement of the $D^{\ast-}$ polarization in the decay $B^0 \to D^{\ast -}\tau^+\nu_{\tau}$, arXiv:1903.03102 [hep-ex].

\bibitem{Tanaka:2012nw} 
M.~Tanaka and R.~Watanabe, New physics in the weak interaction of $\bar B\to D^{(*)}\tau\bar\nu$, Phys.\ Rev.\ D {\bf 87}, 034028 (2013)
\href{http://arxiv.org/abs/1212.1878}{[arXiv:1212.1878 [hep-ph]]}.

\bibitem{Alok:2016qyh} 
A.~K.~Alok, D.~Kumar, S.~Kumbhakar and S.~U.~Sankar, $D^{*}$ polarization as a probe to discriminate new physics in $\bar{B}\to D^{*} \tau \bar{\nu}$, Phys.\ Rev.\ D {\bf 95}, 115038 (2017)
\href{http://arxiv.org/abs/1606.03164}{[arXiv:1606.03164 [hep-ph]]}.


\bibitem{Murgui:2019czp} 
C.~Murgui, A.~Peñuelas, M.~Jung and A.~Pich, Global fit to $b \to c \tau \nu$ transitions, arXiv:1904.09311 [hep-ph].

\bibitem{Bardhan:2019ljo} 
 D.~Bardhan and D.~Ghosh, $B$-meson charged current anomalies: The post-Moriond 
  status,   Phys.\ Rev.\ D {\bf 100}, 011701 (2019)  [arXiv:1904.10432 [hep-ph]].
 
\bibitem{Shi:2019gxi} 
  R.~X.~Shi, L.~S.~Geng, B.~Grinstein, S.~Jäger and J.~Martin Camalich, Revisiting the new-physics interpretation of the $b\to c\tau\nu$ data, arXiv:1905.08498 [hep-ph].
  
\bibitem{Asadi:2019xrc} 
P.~Asadi and D.~Shih, Maximizing the Impact of New Physics in $b\rightarrow c \tau \nu$ Anomalies, arXiv:1905.03311 [hep-ph].

\bibitem{Blanke:2019qrx} 
  M.~Blanke, A.~Crivellin, T.~Kitahara, M.~Moscati, U.~Nierste and I.~Nisandzic, Addendum: "Impact of polarization observables and $B_c\to \tau \nu$ on new physics explanations of the $b\to c \tau \nu$ anomaly", arXiv:1905.08253 [hep-ph].
\bibitem{Alok:2017qsi} 
A.~K.~Alok, D.~Kumar, J.~Kumar, S.~Kumbhakar and S.~U.~Sankar, New physics solutions for $R_D$ and $R_{D^*}$,  JHEP {\bf 1809}, 152 (2018) [arXiv:1710.04127 [hep-ph]].
  
\bibitem{Huang:2018nnq} 
Z.~R.~Huang, Y.~Li, C.~D.~Lu, M.~A.~Paracha and C.~Wang, Footprints of new physics in $b\to c\tau\nu$ transitions, Phys.\ Rev.\ D {\bf 98}, 095018 (2018)   [arXiv:1808.03565 [hep-ph]].
  
\bibitem{Azatov:2018knx} 
A.~Azatov, D.~Bardhan, D.~Ghosh, F.~Sgarlata and E.~Venturini, Anatomy of $b \to c \tau \nu$ anomalies,   JHEP {\bf 1811}, 187 (2018) [arXiv:1805.03209 [hep-ph]].
  
\bibitem{Bhattacharya:2018kig} 
S.~Bhattacharya, S.~Nandi and S.~Kumar Patra, $b \to c \tau \nu_{\tau}$ Decays: A Catalogue to Compare, Constrain, and Correlate New Physics Effects, arXiv:1805.08222 [hep-ph].
  
\bibitem{Jung:2018lfu} 
M.~Jung and D.~M.~Straub, Constraining new physics in $b\to c\ell\nu$ transitions,  JHEP {\bf 1901}, 009 (2019)
[arXiv:1801.01112 [hep-ph]].
  
 
  




\bibitem{Tran:2018kuv} 
C.~T.~Tran, M.~A.~Ivanov, J.~G.~K\"{o}rner and P.~Santorelli, Implications of new physics in the decays $B_c \to (J/\psi,\eta_c)\tau\nu$, Phys.\ Rev.\ D {\bf 97}, 054014 (2018)   \href{http://arxiv.org/abs/1801.06927}{arXiv:1801.06927 [hep-ph]}.

\bibitem{Iguro:2018vqb} 
S.~Iguro, T.~Kitahara, R.~Watanabe and K.~Yamamoto, $D^{\ast}$ polarization vs. $R_{D^{(\ast)}}$ anomalies in the leptoquark models, JHEP {\bf 1902}, 194 (2019) [arXiv:1811.08899 [hep-ph]].

\bibitem{Blanke:2018yud} 
M.~Blanke, A.~Crivellin, S.~de Boer, M.~Moscati, U.~Nierste, I.~Nišandžić and T.~Kitahara, Impact of polarization observables and $ B_c\to \tau \nu$ on new physics explanations of the $b\to c \tau \nu$ anomaly, arXiv:1811.09603 [hep-ph].

\bibitem{Biswas:2018jun} 
A.~Biswas, D.~K.~Ghosh, S.~K.~Patra and A.~Shaw, $b \to c \ell \nu$ anomalies in light of extended scalar sectors,
\href{http://arxiv.org/abs/1801.03375}{arXiv:1801.03375 [hep-ph]}.

\bibitem{Iguro:2018fni} 
S.~Iguro, Y.~Omura and M.~Takeuchi, Test of the $R(D^{(*)})$ anomaly at the LHC, Phys.\ Rev.\ D {\bf 99}, 075013 (2019)   [arXiv:1810.05843 [hep-ph]].

\bibitem{Fraser:2018aqj} 
S.~Fraser, C.~Marzo, L.~Marzola, M.~Raidal and C.~Spethmann, Towards a viable scalar interpretation of $R_{D^{(*)}}$, Phys.\ Rev.\ D {\bf 98}, 035016 (2018) [arXiv:1805.08189 [hep-ph]].
  
\bibitem{Martinez:2018ynq} 
R.~Martinez, C.~F.~Sierra and G.~Valencia, Beyond $\mathcal{R}(D^{(*)})$ with the general type-III 2HDM for $b\to c\tau\nu$, Phys.\ Rev.\ D {\bf 98}, 115012 (2018)  [arXiv:1805.04098 [hep-ph]].

\bibitem{Dhargyal:2016eri} 
L.~Dhargyal, $R(D^{(*)})$ and $\mathcal{B}r(B \rightarrow \tau\nu_{\tau})$ in a Flipped/Lepton-Specific 2HDM with anomalously enhanced charged Higgs coupling to $\tau$/b, Phys.\ Rev.\ D {\bf 93}, 115009 (2016)
  [arXiv:1605.02794 [hep-ph]].
    
\bibitem{Crivellin:2015hha} 
A.~Crivellin, J.~Heeck and P.~Stoffer, A perturbed lepton-specific two-Higgs-doublet model facing experimental hints for physics beyond the Standard Model, Phys. Rev. Lett. {\bf 116}, 081801 (2016). \href{http://arxiv.org/abs/1507.07567}{arXiv:1507.07567 [hep-ph]} 

\bibitem{Iguro:2017ysu} 
S.~Iguro and K.~Tobe, $R(D^{(*)})$ in a general two Higgs doublet model, Nucl.\ Phys.\ B {\bf 925}, 560 (2017) \href{http://arxiv.org/abs/1708.06176}{[arXiv:1708.06176 [hep-ph]]}.

\bibitem{Wei:2017ago} 
M.~Wei and Y.~Chong-Xing, Charged Higgs bosons from the 3-3-1 models and the $\mathcal{R}(D^{(*)})$ anomalies, Phys.\ Rev.\ D {\bf 95}, 035040 (2017).  \href{http://arxiv.org/abs/1702.01255}{[arXiv:1702.01255 [hep-ph]]}
  
\bibitem{Celis:2016azn} 
A.~Celis, M.~Jung, X.~Q.~Li and A.~Pich, Scalar contributions to $b\to c (u) \tau \nu$ transitions, Phys.\ Lett.\ B {\bf 771}, 168 (2017). \href{http://arxiv.org/abs/1612.07757}{[arXiv:1612.07757 [hep-ph]]}

\bibitem{Chen:2017eby} 
C.~H.~Chen and T.~Nomura, Charged-Higgs on $R_{D^{(*)}}$, $\tau$ polarization, and FBA, Eur.\ Phys.\ J.\ C {\bf 77}, 631 (2017). \href{http://arxiv.org/abs/1703.03646}{arXiv:1703.03646 [hep-ph]}  

\bibitem{Sakaki:2012ft} 
Y. Sakaki and H. Tanaka, Constraints on the charged scalar effects using the forward-backward asymmetry on $B \to D^{(*)}\tau \bar{\nu}_\tau$, Phys. Rev. D \textbf{87}, 054002 (2013). \href{http://arxiv.org/abs/1205.4908}{[arXiv:1205.4908 [hep-ph]]}

\bibitem{Crivellin:2012ye} 
A. Crivellin, C. Greub, and A. Kokulu, Explaining $B \to D\tau\nu$, $B \to D^*\tau\nu$ and $B \to\tau\nu$ in a 2HDM of type III, Phys. Rev. D \textbf{86}, 054014 (2012) 
\href{http://arxiv.org/abs/1206.2634}{[arXiv:1206.2634]}.

\bibitem{Crivellin:2013wna} 
A. Crivellin, C. Greub, and A. Kokulu, Flavor-phenomenology of two-Higgs-doublet models with generic Yukawa structure, Phys. Rev. D \textbf{87}, 094031 (2013). \href{http://arxiv.org/abs/1303.5877}{[arXiv:1303.5877 [hep-ph]]}

\bibitem{Celis:2012dk}
A. Celis, M. Jung, X.-Q. Li, and A. Pich, Sensitivity to charged scalars in $B \to D^{(*)} \tau \nu_\tau$ and $B \to \tau \nu_\tau$ decays, JHEP \textbf{01}, 054 (2013). \href{http://arxiv.org/abs/1210.8443}{[arXiv:1210.8443 [hep-ph]]}

\bibitem{Ko:2012sv}
P. Ko, Y. Omura, and C. Yu, $B \to D^{(*)} \tau\nu$ and $B \to \tau\nu$ in chiral $U(1)^\prime$ models with flavored multi Higgs doublets, JHEP \textbf{03}, 151 (2013). \href{http://arxiv.org/abs/1212.4607}{[arXiv:1212.4607 [hep-ph]]}

\bibitem{Ko:2017lzd} 
P.~Ko, Y.~Omura, Y.~Shigekami and C.~Yu, The LHCb anomaly and $B$ physics in flavored Z$^{\prime}$ models with flavored Higgs doublets, Phys. Rev. D \textbf{95}, 115040 (2017). \href{http://arxiv.org/abs/1702.08666}{[arXiv:1702.08666 [hep-ph]]} 

\bibitem{Li:2018rax} 
  S.~P.~Li, X.~Q.~Li, Y.~D.~Yang and X.~Zhang, ${R}_{D^{\left(*\right)}},{R}_{K^{\left(*\right)}}$ and neutrino mass in the 2HDM-III with right-handed neutrinos,  JHEP {\bf 1809}, 149 (2018)  [arXiv:1807.08530 [hep-ph]]. 
  


\bibitem{Assad:2017iib} 
N.~Assad, B.~Fornal and B.~Grinstein, Baryon Number and Lepton Universality Violation in Leptoquark and Diquark Models, Phys.\ Lett.\ B {\bf 777}, 324 (2018) [arXiv:1708.06350 [hep-ph]].

\bibitem{Hati:2019ufv} 
C.~Hati, J.~Kriewald, J.~Orloff and A.~M.~Teixeira, A nonunitary interpretation for a single vector leptoquark combined explanation to the $B$-decay anomalies,
arXiv:1907.05511 [hep-ph].
 
\bibitem{Hati:2018fzc} 
  C.~Hati, G.~Kumar, J.~Orloff and A.~M.~Teixeira, Reconciling $B$-meson decay anomalies with neutrino masses, dark matter and constraints from flavour violation, JHEP {\bf 1811}, 011 (2018) [arXiv:1806.10146 [hep-ph]].

\bibitem{Fornal:2018dqn} 
B.~Fornal, S.~A.~Gadam and B.~Grinstein, Left-Right SU(4) Vector Leptoquark Model for Flavor Anomalies, Phys.\ Rev.\ D {\bf 99}, no. 5, 055025 (2019) [arXiv:1812.01603 [hep-ph]].
  
\bibitem{Yan:2019hpm}
H.~Yan, Y.~D.~Yang and X.~B.~Yuan, Phenomenology of $b\to c\tau\bar\nu$ decays in a scalar leptoquark model, Chin. Phys. C 43, (2019) 083105 [arXiv:1905.01795 [hep-ph]].
  
\bibitem{Cornella:2019hct} 
C.~Cornella, J.~Fuentes-Martin and G.~Isidori, Revisiting the vector leptoquark explanation of the $B$-physics anomalies, arXiv:1903.11517 [hep-ph].
  
\bibitem{Becirevic:2016yqi} 
D.~Becirevic, S.~Fajfer, N.~Kosnik and O.~Sumensari, Leptoquark model to explain the $B$-physics anomalies, $R_K$ and $R_D$,  Phys.\ Rev.\ D {\bf 94}, 115021 (2016)  [arXiv:1608.08501 [hep-ph]].

\bibitem{Becirevic:2018afm} 
  D.~Becirevic, I.~Dorsner, S.~Fajfer, N.~Kosnik, D.~A.~Faroughy and O.~Sumensari, Scalar leptoquarks from grand unified theories to accommodate the $B$-physics anomalies, Phys.\ Rev.\ D {\bf 98}, 055003 (2018)
  [arXiv:1806.05689 [hep-ph]].

\bibitem{Alonso:2015sja} 
  R.~Alonso, B.~Grinstein and J.~Martin Camalich, Lepton universality violation and lepton flavor conservation in $B$-meson decays,  JHEP {\bf 1510}, 184 (2015)   [arXiv:1505.05164 [hep-ph]].

  \bibitem{Calibbi:2015kma} 
  L.~Calibbi, A.~Crivellin and T.~Ota, Effective Field Theory Approach to $b \to s \ell \ell^{(\prime)}$, $B \to K^{(*)} \nu\bar{\nu}$ and $B \to D^{(*)}\ell \nu$ with Third Generation Couplings, Phys.\ Rev.\ Lett.\  {\bf 115}, 181801 (2015)  [arXiv:1506.02661 [hep-ph]].
  
  \bibitem{Fajfer:2015ycq} 
  S.~Fajfer and N.~Kosnik, Vector leptoquark resolution of $R_K$ and $R_{D^{(*)}}$ puzzles, 
  Phys.\ Lett.\ B {\bf 755}, 270 (2016)   [arXiv:1511.06024 [hep-ph]].

 \bibitem{Barbieri:2015yvd} 
  R.~Barbieri, G.~Isidori, A.~Pattori and F.~Senia, Anomalies in $B$-decays and $U(2)$ flavour symmetry,
  Eur.\ Phys.\ J.\ C {\bf 76}, 67 (2016)   [arXiv:1512.01560 [hep-ph]].

  \bibitem{Barbieri:2016las} 
  R.~Barbieri, C.~W.~Murphy and F.~Senia, $B$-decay Anomalies in a Composite Leptoquark Model,
  Eur.\ Phys.\ J.\ C {\bf 77}, 8 (2017)   [arXiv:1611.04930 [hep-ph]].
  
  \bibitem{Hiller:2016kry} 
  G.~Hiller, D.~Loose and K.~Sch\"{o}nwald, Leptoquark Flavor Patterns \& $B$ Decay Anomalies, JHEP {\bf 1612}, 027 (2016). \href{http://arxiv.org/abs/1609.08895}{[arXiv:1609.08895 [hep-ph]]}
  
  \bibitem{Bhattacharya:2016mcc} 
  B.~Bhattacharya, A.~Datta, J.~P.~Gu\'{e}vin, D.~London and R.~Watanabe, Simultaneous Explanation of the $R_K$ and $R_{D^{(*)}}$ Puzzles: a Model Analysis,  JHEP {\bf 1701}, 015 (2017)   [arXiv:1609.09078 [hep-ph]].

  \bibitem{Buttazzo:2017ixm} 
  D.~Buttazzo, A.~Greljo, G.~Isidori and D.~Marzocca, $B$-physics anomalies: a guide to combined explanations,
  JHEP {\bf 1711}, 044 (2017)   [arXiv:1706.07808 [hep-ph]].
  
  \bibitem{Kumar:2018kmr} 
  J.~Kumar, D.~London and R.~Watanabe, Combined Explanations of the $b \to s \mu^+ \mu^-$ and $b \to c \tau^- {\bar\nu}$ Anomalies: a General Model Analysis, Phys.\ Rev.\ D {\bf 99}, no. 1, 015007 (2019)   [arXiv:1806.07403 [hep-ph]].

 
 \bibitem{Assad:2017iib} 
  N.~Assad, B.~Fornal and B.~Grinstein, Baryon Number and Lepton Universality Violation in Leptoquark and Diquark Models, Phys.\ Lett.\ B {\bf 777}, 324 (2018)   [arXiv:1708.06350 [hep-ph]]. 

  \bibitem{DiLuzio:2017vat} 
  L.~Di Luzio, A.~Greljo and M.~Nardecchia, Gauge leptoquark as the origin of B-physics anomalies, 
  Phys.\ Rev.\ D {\bf 96}, no. 11, 115011 (2017)   [arXiv:1708.08450 [hep-ph]].
  
  \bibitem{Calibbi:2017qbu} 
  L.~Calibbi, A.~Crivellin and T.~Li, Model of vector leptoquarks in view of the $B$-physics anomalies,
  Phys.\ Rev.\ D {\bf 98}, no. 11, 115002 (2018)   [arXiv:1709.00692 [hep-ph]].

  \bibitem{Barbieri:2017tuq} 
  R.~Barbieri and A.~Tesi, $B$-decay anomalies in Pati-Salam SU(4),  Eur.\ Phys.\ J.\ C {\bf 78}, no. 3, 193 (2018)   [arXiv:1712.06844 [hep-ph]].

  \bibitem{Blanke:2018sro} 
  M.~Blanke and A.~Crivellin, $B$ Meson Anomalies in a Pati-Salam Model within the Randall-Sundrum Background,
  Phys.\ Rev.\ Lett.\  {\bf 121}, no. 1, 011801 (2018)  [arXiv:1801.07256 [hep-ph]].

  \bibitem{Greljo:2018tuh} 
  A.~Greljo and B.~A.~Stefanek, Third family quark–lepton unification at the TeV scale, Phys.\ Lett.\ B {\bf 782}, 131 (2018)   [arXiv:1802.04274 [hep-ph]].
  
  \bibitem{Bauer:2015knc} 
  M.~Bauer and M.~Neubert, Minimal Leptoquark Explanation for the R$_{D^{(*)}}$ , R$_K$ , and $(g-2)_g$ Anomalies,  Phys.\ Rev.\ Lett.\  {\bf 116}, 141802 (2016) \href{http://arxiv.org/abs/1511.01900}{arXiv:1511.01900 [hep-ph]}. 
  
  \bibitem{Bordone:2018nbg} 
  M.~Bordone, C.~Cornella, J.~Fuentes-Martín and G.~Isidori, Low-energy signatures of the $\mathrm{PS}^3$ model: from $B$-physics anomalies to LFV,  JHEP {\bf 1810}, 148 (2018)   [arXiv:1805.09328 [hep-ph]].

  
  \bibitem{Crivellin:2018yvo} 
  A.~Crivellin, C.~Greub, D.~Müller and F.~Saturnino, Importance of Loop Effects in Explaining the Accumulated Evidence for New Physics in B Decays with a Vector Leptoquark, Phys.\ Rev.\ Lett.\  {\bf 122}, no. 1, 011805 (2019)  [arXiv:1807.02068 [hep-ph]].

  \bibitem{DiLuzio:2018zxy} 
  L.~Di Luzio, J.~Fuentes-Martin, A.~Greljo, M.~Nardecchia and S.~Renner, Maximal Flavour Violation: a Cabibbo mechanism for leptoquarks,  JHEP {\bf 1811}, 081 (2018)   [arXiv:1808.00942 [hep-ph]].

  \bibitem{Cai:2017wry} 
Y.~Cai, J.~Gargalionis, M.~A.~Schmidt and R.~R.~Volkas, Reconsidering the One Leptoquark solution: flavor anomalies and neutrino mass,   JHEP {\bf 1710}, 047 (2017) \href{http://arxiv.org/abs/1704.05849}{[arXiv:1704.05849 [hep-ph]]}.
  
\bibitem{Crivellin:2017zlb} 
A.~Crivellin, D.~M\"{u}ller and T.~Ota, Simultaneous Explanation of $R(D^{(*)})$ and $b\to s\mu^+\mu^-$: The Last Scalar Leptoquarks Standing, JHEP {\bf 1709}, 040 (2017). \href{http://arxiv.org/abs/1703.09226}{[arXiv:1703.09226 [hep-ph]]} 
  
 \bibitem{Li:2016vvp} 
X.~Q.~Li, Y.~D.~Yang and X.~Zhang, Revisiting the one leptoquark solution to the $R(D^{(*)})$ anomalies and its phenomenological implications,  JHEP {\bf 1608}, 054 (2016).  \href{http://arxiv.org/abs/1605.09308}{[arXiv:1605.09308 [hep-ph]]}
  
  \bibitem{Das:2016vkr} 
D.~Das, C.~Hati, G.~Kumar and N.~Mahajan, Towards a unified explanation of $R_{D^{(\ast)}}$, $R_{K}$ and $(g-2)_{\mu}$ anomalies in a left-right model with leptoquarks, Phys.\ Rev.\ D {\bf 94}, 055034 (2016). \href{http://arxiv.org/abs/1605.06313}{[arXiv:1605.06313 [hep-ph]]} 
   
\bibitem{Faroughy:2016osc} 
D.~A.~Faroughy, A.~Greljo and J.~F.~Kamenik, Confronting lepton flavor universality violation in B decays with high-$p_T$ tau lepton searches at LHC, Phys.\ Lett.\ B {\bf 764}, 126 (2017). \href{http://arxiv.org/abs/1609.07138}{[arXiv:1609.07138 [hep-ph]]} 
  
\bibitem{Sahoo:2016pet} 
S.~Sahoo, R.~Mohanta and A.~K.~Giri, Explaining the $R_{K}$ and $R_{D^{(*)}}$ anomalies with vector leptoquarks, Phys.\ Rev.\ D {\bf 95}, 035027 (2017)  \href{http://arxiv.org/abs/1609.04367}{[arXiv:1609.04367 [hep-ph]]}. 


\bibitem{Chen:2017hir} 
  C.~H.~Chen, T.~Nomura and H.~Okada, Excesses of muon $g-2$, $R_{D^{(\ast)}}$, and $R_K$ in a leptoquark model, Phys.\ Lett.\ B {\bf 774}, 456 (2017) \href{http://arxiv.org/abs/1703.03251}{[arXiv:1703.03251 [hep-ph]]}

\bibitem{Sakaki:2013bfa} 
Y.~Sakaki, M.~Tanaka, A.~Tayduganov, and R.~Watanabe, Testing leptoquark models in $\bar B \to D^{(*)} \tau\bar\nu$, Phys. Rev. D \textbf{88}, 094012 (2013). \href{http://arxiv.org/abs/1309.0301}{[arXiv:1309.0301 [hep-ph]]}



\bibitem{Dasgupta:2018nzt} 
  S.~Dasgupta, U.~K.~Dey, T.~Jha and T.~S.~Ray, Status of a flavor-maximal nonminimal universal extra dimension model, Phys.\ Rev.\ D {\bf 98}, 055006 (2018)   [arXiv:1801.09722 [hep-ph]].

  
\bibitem{He:2012zp} 
X.~G.~He and G.~Valencia, $B$ decays with $\tau$ leptons in nonuniversal left-right models, Phys.\ Rev.\ D {\bf 87}, 014014 (2013)   [arXiv:1211.0348 [hep-ph]].

\bibitem{He:2017bft}
X.-G. He and G. Valencia, Lepton universality violation and right-handed currents in $b \to c \tau \nu$,  Phys.\ Lett.\ B {\bf 779}, 52 (2018) \href{http://arxiv.org/abs/1711.09525}{arXiv:1711.09525  [hep-ph]}.

\bibitem{Boucenna:2016qad} 
S.~M.~Boucenna, A.~Celis, J.~Fuentes-Martin, A.~Vicente and J.~Virto, Phenomenology of an $SU(2) \times SU(2) \times U(1)$ model with lepton-flavour non-universality, JHEP {\bf 1612}, 059 (2016)  \href{http://arxiv.org/abs/1608.01349}{[arXiv:1608.01349 [hep-ph]]}.

\bibitem{Boucenna:2016wpr}
S.~M.~Boucenna, A.~Celis, J.~Fuentes-Martin, A.~Vicente and J.~Virto, Non-abelian gauge extensions for B-decay anomalies, Phys.\ Lett.\ B {\bf 760}, 214 (2016) \href{http://arxiv.org/abs/1604.03088}{[arXiv:1604.03088 [hep-ph]]}.

\bibitem{Greljo:2015mma} 
A.~Greljo, G.~Isidori and D.~Marzocca, On the breaking of Lepton Flavor Universality in B decays,  JHEP {\bf 1507}, 142 (2015)   [arXiv:1506.01705 [hep-ph]].

\bibitem{Abdullah:2018ets} 
M.~Abdullah, J.~Calle, B.~Dutta, A.~Fl\'{o}rez and D.~Restrepo, Probing a simplified $W^{\prime}$ model of $R(D^{(\ast)})$ anomalies using $b$-tags, $\tau$ leptons and missing energy, Phys.\ Rev.\ D {\bf 98}, 055016 (2018) [arXiv:1805.01869 [hep-ph]].

\bibitem{Greljo:2018tzh} 
 A.~Greljo, J.~Martin Camalich and J.~D.~Ruiz-\'{A}lvarez, Mono-$\tau$ Signatures at the LHC Constrain Explanations of B-decay Anomalies, Phys.\ Rev.\ Lett.\  {\bf 122}, 131803 (2019) [arXiv:1811.07920 [hep-ph]].
 
\bibitem{Carena:2018cow} 
  M.~Carena, E.~Meg\'{i}as, M.~Qu\'{i}ros and C.~Wagner, $ {R}_{D^{\left(*\right)}} $ in custodial warped space,  JHEP {\bf 1812}, 043 (2018) [arXiv:1809.01107 [hep-ph]].

\bibitem{Babu:2018vrl} 
K.~S.~Babu, R.~N.~Mohapatra and B.~Dutta, A Theory of $R(D^*,D)$ Anomaly with Right-Handed Currents,   JHEP {\bf 1901}, 168 (2019) [arXiv:1811.04496 [hep-ph]].
  
\bibitem{Asadi:2018wea}  
  P.~Asadi, M.~R.~Buckley and D.~Shih, It’s all right(-handed neutrinos): a new W$^{\prime}$ model for the $ {R}_{D^{{\left(\ast \right)}}} $ anomaly, JHEP {\bf 1809}, 010 (2018) [arXiv:1804.04135 [hep-ph]].  

\bibitem{Greljo:2018ogz} 
  A.~Greljo, D.~J.~Robinson, B.~Shakya and J.~Zupan, R(D$^{(\ast)}$) from $W^{\prime}$ and right-handed neutrinos, JHEP {\bf 1809}, 169 (2018)  [arXiv:1804.04642 [hep-ph]].

\bibitem{Robinson:2018gza} 
D.~J.~Robinson, B.~Shakya and J.~Zupan, Right-handed Neutrinos and $R(D^{(*)})$, JHEP {\bf 1902}, 119 (2019) [arXiv:1807.04753 [hep-ph]].

\bibitem{Asadi:2018sym} 
P.~Asadi, M.~R.~Buckley and D.~Shih, Asymmetry Observables and the Origin of $R_{D^{(*)}}$ Anomalies, Phys.\ Rev.\ D {\bf 99}, 035015 (2019) arXiv:1810.06597 [hep-ph].
  

\bibitem{Azatov:2018kzb} 
A.~Azatov, D.~Barducci, D.~Ghosh, D.~Marzocca and L.~Ubaldi, Combined explanations of B-physics anomalies: the sterile neutrino solution, JHEP {\bf 1810}, 092 (2018)  [arXiv:1807.10745 [hep-ph]].

\bibitem{Cvetic:2017gkt} 
G.~Cveti\v{c}, F.~Halzen, C.~S.~Kim and S.~Oh,  Anomalies in (semi)-leptonic $B$ Decays $B^{\pm} \to \tau^{\pm} \nu$, $B^{\pm} \to D \tau^{\pm} \nu$ and $B^{\pm} \to D^* \tau^{\pm} \nu$, and possible resolution with sterile neutrino, Chin.\ Phys.\ C {\bf 41}, 113102 (2017). \href{http://arxiv.org/abs/1702.04335}{[arXiv:1702.04335 [hep-ph]]}

\bibitem{Altmannshofer:2017poe} 
W.~Altmannshofer, P.~S.~B.~Dev and A.~Soni, $R_{D^{(*)}}$ anomaly: A possible hint for natural supersymmetry with $R$-parity violation, Phys. Rev. D \textbf{96}, 095010 (2017). \href{http://arxiv.org/abs/1704.06659}{[arXiv:1704.06659 [hep-ph]]}
  
\bibitem{Deshpande:2016cpw} 
N.~G.~Deshpande and X.~G.~He, Consequences of R-parity violating interactions for anomalies in $\bar B\to D^{(*)} \tau \bar \nu$ and $b\to s \mu^+\mu^-$, Eur.\ Phys.\ J.\ C {\bf 77}, 134 (2017).
\href{http://arxiv.org/abs/1608.04817}{[arXiv:1608.04817 [hep-ph]]}
  
\bibitem{Deshpande:2012rr}
N. G. Deshpande and A. Menon, Hints of R-parity violation in $B$ decays into $\tau\nu$, JHEP \textbf{01}, 025 (2013).
\href{http://arxiv.org/abs/1208.4134}{[arXiv:1208.4134 [hep-ph]]}

\bibitem{Zhu:2016xdg} 
J.~Zhu, H.~M.~Gan, R.~M.~Wang, Y.~Y.~Fan, Q.~Chang and Y.~G.~Xu, Probing the R-parity violating supersymmetric effects in the exclusive $b\to c\ell^-\bar{\nu}_\ell$ decays, Phys.\ Rev.\ D {\bf 93}, 094023 (2016). \href{http://arxiv.org/abs/1602.06491}{[arXiv:1602.06491 [hep-ph]]} 
  
\bibitem{Wei:2018vmk} 
B.~Wei, J.~Zhu, J.~H.~Shen, R.~M.~Wang and G.~R.~Lu, Probing the $R$-parity violating supersymmetric effects in $B_c\to J/\psi\ell^-\bar{\nu}_{\ell},\eta_c\ell^-\bar{\nu}_{\ell}$ and $\Lambda_b\to\Lambda_c\ell^-\bar{\nu}_{\ell}$ decays, Nucl.\ Phys.\ B {\bf 934}, 380 (2018)  \href{http://arxiv.org/abs/1801.00917}{[arXiv:1801.00917 [hep-ph]]}.

\bibitem{Hu:2018lmk}
Q.~Y.~Hu, X.~Q.~Li, Y.~Muramatsu and Y.~D.~Yang, $R$-parity violating solutions to the $R_{D^{(\ast)}}$ anomaly and their GUT-scale unifications, Phys.\ Rev.\ D {\bf 99}, 015008 (2019) [arXiv:1808.01419 [hep-ph]].

\bibitem{Trifinopoulos:2018rna} 
S.~Trifinopoulos, Revisiting $R$-parity violating interactions as an explanation of the B-physics anomalies, Eur.\ Phys.\ J.\ C {\bf 78}, 803 (2018) [arXiv:1807.01638 [hep-ph]].
  
\bibitem{Alonso:2016oyd} 
R.~Alonso, B.~Grinstein and J.~Martin Camalich, Lifetime of $B_c^-$ Constrains Explanations for Anomalies in  $B\to D^{(*)}\tau\nu$, Phys.\ Rev.\ Lett.\  {\bf 118}, 081802 (2017). \href{http://arxiv.org/abs/1611.06676}{[arXiv:1611.06676 [hep-ph]]}   

\bibitem{Akeroyd:2017mhr}
A.~G.~Akeroyd and C.~H.~Chen, Constraint on the branching ratio of $B_c \to \tau \nu$ from LEP1 and consequences for R(D(*)) anomaly, Phys. Rev. D \textbf{96}, 075011 (2017). \href{http://arxiv.org/abs/1708.04072}{[arXiv:1708.04072 [hep-ph]]}.

\bibitem{Tanabashi:2018oca} 
M.~Tanabashi {\it et al.} (Particle Data Group), Review of Particle Physics, Phys.\ Rev.\ D {\bf 98}, 030001 (2018) \href{http://pdg.lbl.gov}{\texttt{[http://pdg.lbl.gov]}}.

\bibitem{Colquhoun:2015oha} 
B.~Colquhoun {\it et al.} (HPQCD Collaboration), B-meson decay constants: a more complete picture from full lattice QCD, Phys.\ Rev.\ D {\bf 91}, 114509 (2015). 
\href{http://arxiv.org/abs/1503.05762}{\texttt{[arXiv:1503.05762 [hep-lat]]}}.

\bibitem{Kamali:2018bdp} 
S.~Kamali, New physics in inclusive semileptonic $B$ decays including nonperturbative corrections, Int.\ J.\ Mod.\ Phys.\ A {\bf 34}, 1950036 (2019) [arXiv:1811.07393 [hep-ph]].

\bibitem{Feruglio:2017rjo} 
F.~Feruglio, P.~Paradisi and A.~Pattori, On the Importance of Electroweak Corrections for B Anomalies, JHEP {\bf 1709}, 061 (2017)  [arXiv:1705.00929 [hep-ph]]

\bibitem{Aaboud:2018vgh} 
M.~Aaboud {\it et al.} [ATLAS Collaboration], Search for High-Mass Resonances Decaying to $\tau\nu$ in pp Collisions at $\sqrt{s}$=13  TeV with the ATLAS Detector, Phys.\ Rev.\ Lett.\  {\bf 120}, 161802 (2018) [arXiv:1801.06992 [hep-ex]].

\bibitem{Sirunyan:2018lbg} 
  A.~M.~Sirunyan {\it et al.} [CMS Collaboration], Search for a W' boson decaying to a $\tau$ lepton and a neutrino in proton-proton collisions at $\sqrt{s} =$ 13 TeV, Phys.\ Lett.\ B {\bf 792}, 107 (2019) [arXiv:1807.11421 [hep-ex]].

  
\bibitem{Bhattacharya:2019olg} 
B.~Bhattacharya, A.~Datta, S.~Kamali and D.~London, CP Violation in ${\bar B}^0\to D^{*+}\mu^-{\bar\nu}_\mu$, JHEP {\bf 1905}, 191 (2019)   [arXiv:1903.02567 [hep-ph]].

  
\end{thebibliography}


\end{document}